\documentclass[aps,pre,twocolumn,showpacs,floatfix]{revtex4-1}
\usepackage{graphicx}
\usepackage{graphics}
\usepackage{amssymb}
\usepackage{hyperref}
\usepackage{latexsym}
\usepackage{amsmath}
\usepackage{textcomp}
\usepackage{amsfonts}
\usepackage{color}
\usepackage{bm}

\newcommand{\bkappa}{\bm{\kappa}}
\newcommand{\Dmat}{\tilde{\mathcal D}}
\newcommand{\vBZ}{v_{\rm BZ}}
\newcommand{\ed}{e_u}
\newcommand{\bR}{{\bf R}}
\newcommand{\barR}{\bar{\bf R}}
\newcommand{\barchi}{\bar{\chi}}
\newcommand{\Emat}{\mathsf E}

\begin{document}

\title{Non-affine displacements in crystalline solids in the harmonic limit}



\author{Saswati Ganguly$^1$, Surajit Sengupta$^{2,1}$, Peter Sollich$^{3}$, Madan Rao$^{4,5}$ 
}
\affiliation{$^1$ Indian Association for the Cultivation of Science, 2A\&2B Raja S. C. Mullick Road, Jadavpur, Kolkata 700032, India \\
$^2$ TIFR Centre for Interdisciplinary Sciences, 21, Brundavan Colony, Narsingi, Hyderabad 500075, India \\
$^3$ King's College London, Department of Mathematics, Strand, London WC2R 2LS, U.K.\\ 
$^4$ Raman Research Institute,C.V. Raman Avenue, Bangalore 560080, India, \\
$^5$ National Centre for Biological Sciences (TIFR), Bellary Road, Bangalore 560 065, India
}


%
%
%

\begin{abstract}
A systematic coarse graining of microscopic atomic displacements generates a local elastic deformation tensor ${\mathsf D}$ as well as a positive definite scalar $\chi$ measuring {\it non-affinity}, i.e.\ the extent to which the displacements are not representable as affine deformations of a reference crystal. We perform an exact calculation of  the statistics of $\chi$ and ${\mathsf D}$ and their spatial correlations for solids at low temperatures, within a harmonic approximation and in one and two dimensions. We 
obtain the joint distribution $P(\chi, {\mathsf D})$ and the two point spatial correlation functions for $\chi$ and ${\mathsf D}$. We show that non-affine and affine deformations are coupled even in a harmonic solid, with a strength that depends on the size of the coarse graining volume $\Omega$ and dimensionality. 
As a corollary to our work, we identify the field, $h_{\chi}$, conjugate to $\chi$ and show that this field may be tuned to produce a transition to a state where the ensemble average, $\langle \chi \rangle$, and the correlation length of $\chi$ diverge. Our work should be useful as a template for understanding non-affine displacements in realistic systems with or without disorder and as a means for developing computational tools for studying the effects of non-affine displacements in melting, plastic flow and the glass transition.    
\end{abstract}

\maketitle

\section{Introduction}
\label{intro}
Understanding the mechanical response of soft and disordered solids~\cite{cates-evans} such as polymer gels~\cite{polymer}, fabric~\cite{fabric}, foams~\cite{foams}, colloids~\cite{colloids}, granular matter~\cite{granular} and glasses~\cite{falk, falk-review} is challenging because one is often lead to questions that lie on the boundaries of classical elasticity theory~\cite{LL, CL, MTE}. For example, under external stress, particles $i$ within a solid undergo displacements ${\bf u}_i = {\bf r}_i - {\bf R}_i$ away from some chosen reference configuration ${\bf R}_i$ to their displaced positions ${\bf r}_i$. In a conventional homogeneous solid, such displacements are {\em affine}, in the sense that they can be expressed as ${\bf u}_i = {\mathsf D} {\bf R}_i$, where  ${\mathsf D} = {\mathsf K}^{-1} {\mathsf \sigma}$ is the deformation tensor related to the external stress ${\mathsf \sigma}$ via the tensor of elastic constants ${\mathsf K}$. This is not true if the solid is disordered at a microscopic level.  

One of the principal sources of {\em non-affinity}  is a space (and possibly even time) dependent elastic constant~\cite{zero-T}. The local environment in a disordered solid varies in space, depending crucially on local connectivity or coordination such that the local displacement ${\bf u}_i$ may not be simply related to the applied stress $\sigma$. Such non-affine displacements are present even at zero temperature, are material dependent, and vanish only for homogeneous crystalline media without defects. 

In this paper we explore another, perhaps complementary, source of non-affinity, namely that which arises due to thermal fluctuations and coarse graining. Elastic properties of materials emerge upon coarse graining microscopic particle displacements~\cite{models, coarse-graining, elast-breakdown, zahn, kers1,kers2, zhang} over a coarse-graining volume $\Omega$. A systematic finite size scaling analysis of the $\Omega$ dependent elastic constants then yields the material properties in the thermodynamic limit~\cite{kers1,kers2}. Such a coarse graining procedure has been used to obtain elastic constants of soft colloidal crystals from video microscopy~\cite{zahn, kers1, zhang} as well as in model solids~\cite{models,kers2}. For distances smaller than the size of $\Omega$, particle displacements, in $d$ dimensions, are necessarily {\em non-affine} since the local distortion ${\mathsf D}$ of ${\Omega}$ is obtained by projecting the displacements of all $N$ particles in $\Omega$ into the $d \times d$-dimensional space of affine distortions which, in general, is smaller than the full $Nd$-dimensional configuration space available. The generation of non-affinity $\chi$, defined as the sum of squares of all the particle displacements which do not belong to the projected space of affine distortions, is therefore a necessary consequence of the coarse-graining procedure. Here we take a detailed look at this process and present an exact calculation for the probability distributions and correlation functions for $\chi$ and ${\mathsf D}$ for harmonic solids in $d = 1$ and $2$ in the canonical ensemble. Our work allows us to identify the field conjugate to $\chi$, viz.\ $h_{\chi}$, and we show that by tuning this $h_{\chi}$ one may enhance non-affine fluctuations and cause a transition. At this transition all the moments of the probability distribution of $\chi$ diverge, thereby disordering the solid {\em isothermally}. 

There are several reasons why we believe that our work may be useful. Firstly, the harmonic crystal is often the starting point for more realistic calculations of the elastic properties of solids and constitutes an ideal system to which simulation and experimental results~\cite{harm-colloid} can be compared in order to quantify purely anharmonic effects. Secondly, a coarse-grained theory for the mechanical properties of soft solids should contain both the elastic and non-elastic fields ${\mathsf D}$ and $\chi$: our work may provide a hint on how such a theory may be constructed. Thirdly, we believe that it may be possible to extend our calculations to systems with isolated defects or randomness, thus extending the analysis of Ref.~\cite{zero-T} to non-zero temperatures. Finally, our calculations may be used to devise new simulational strategies for understanding the influence of non-affine fluctuations on the mechanical properties of both crystals and glasses and, perhaps, shed more light on the nature of the glass transition itself. 

The paper is organized as follows. In section~\ref{sec:1} we set up the calculation and define $\chi$, ${\mathsf D}$ and the coarse-graining process. In section~\ref{sec:2} we present our calculation for the single point probability distributions for $\chi$ and ${\mathsf D}$ for $d=1$ harmonic chains and the $d=2$ triangular harmonic lattices. Approximate experimental realisations of these systems correspond to mercury chain salts~\cite{Hg} and the spectrin network in red blood corpuscles~\cite{rbc} respectively. In section~\ref{sec:3}, we evaluate the spatial correlation functions for ${\mathsf D}$ and $\chi$. This is followed by a calculation of linear response and identification of the non-affine field in section~\ref{sec:4}. Finally we discuss our results and conclude by giving indications of future directions in section~\ref{sec:5}.
     
\section{Coarse graining and the non-affine parameter}
\label{sec:1}

Consider a neighborhood, $\Omega$, in a $d$ dimensional lattice consisting of $N$ particles $i$ arranged around the central particle $0$ within a cut-off distance $R_\Omega$. Mostly we set $R_\Omega$ equal to the nearest neighbour distance so that $\Omega$ contains all nearest neighbours of particle 0, but in section~\ref{sec:5} we also consider larger $R_\Omega$.
The zero temperature lattice positions that we choose as our reference are ${\bf R}_{i}$ and ${\bf R}_{0}$ and the fluctuating atom positions will be denoted ${\bf r}_{i}$ and ${\bf r}_0$\,\cite{metric}. Define the particle displacements ${\bf u}_i = {\bf r}_{i} - {\bf R}_{i}$, and 
${\bf \Delta}_{i} = {\bf u}_i-{\bf u}_0 = {\bf r}_{i} - {\bf r}_{0} -({\bf R}_{i} - {\bf R}_{0})$ as the displacement of particle $i$ relative to particle 0. We will often use the Fourier transform of the particle displacement,  ${\bf u}_{\bf q}$, which is defined such that the real-space displacements are ${\bf \Delta}_i = {\bf u}_i - {\bf u}_0 = l\, \vBZ^{-1}\int d{\bf q}\, {\bf u}_{\bf q}(e^{i {\bf q \cdot  R}_{i}} - e^{i {\bf q \cdot R}_0})$. Here $l$ is the lattice parameter  and $\vBZ$ is the volume of the Brillouin zone over which the ${\bf q}$ integral is performed.

If the local particle displacements are fully affine then one has ${\bf u}_i = {\mathsf D} {\bf R}_i$, and hence 
${\bf \Delta}_{i} = {\mathsf D}({\bf R}_{i} - {\bf R}_{0})$. Generically the displacements will contain a non-affine component and the coarse-grained local deformation tensor ${\mathsf D}$ can then be defined~\cite{falk,bagi}  as the one that minimizes $\sum_i
[{\bf \Delta}_{i} - {\mathsf D}({\bf R}_{i} - {\bf R}_{0})]^2$. The minimal value of this quantity is the non-affinity parameter $\chi$. 

To simplify the notation we arrange the $Nd$ relative displacement components
$\Delta_{i \alpha}$, where the index $\alpha = 1,\dots,d$ labels the spatial components, into an $Nd$-dimensional vector ${\bf \Delta}$.
We similarly define a vector ${\bf e}$ whose components are the $d^2$
elements of the local deformation tensor, ${\mathsf
D}_{\alpha\gamma}$, arranged as a linear array (viz.\ ${\bf e} =
(D_{11}, D_{12}, \dots , D_{1d}, D_{21}, \dots , D_{Nd})$), and a
matrix ${\mathsf R}$ of size $Nd\times d^{2}$ with elements ${\mathsf
R}_{i\alpha,\gamma\gamma^{\prime}} =
\delta_{\alpha\gamma}(R_{i\gamma^{\prime}}-R_{0\gamma^{\prime}})$
where the $R_{i\gamma^{\prime}}$ and $R_{0,\gamma^{\prime}}$ are the
components of the lattice positions ${\bf R}_{i}$ and ${\bf R}_{i}$,
respectively. Below we use the notations ${\bf e}$ and ${\mathsf D}$ for the deformation tensor interchangeably, as convenient in the context. 

As explained above, we will define the non-affinity parameter $\chi$ as the residual sum of squares of all the displacements of the particles in $\Omega$ after fitting the best affine deformation, measured with respect to the reference configuration~\cite{falk}. The local deformation $\bf e$ is thus obtained by minimising the positive definite quantity $(\Delta-{\mathsf R}{\bf e})^2$ with respect to $\bf e$:
\begin{eqnarray}
\chi&=&{\rm min}_{\bf e}\left({\bf \Delta}-{\mathsf R}{\bf e}\right)^{2} \nonumber \\
&=&{\rm min}_{\bf e}({\bf \Delta}^{t}{\bf \Delta}-{\bf \Delta}^{t}{\mathsf R}{\bf e}-{\bf e}^{t}{\mathsf R}^{t}{\bf \Delta}+{\bf e}^{t}{\mathsf R}^{t}{\mathsf R}{\bf e})\label{chidef}
\end{eqnarray}
Here the superscript $t$ denotes the transpose operation. The coarse-grained local deformation, i.e.\ the value of ${\bf e}$ where the minimum is obtained, can then be written as
\begin{equation}
{\bf e} = {\mathsf Q}{\bf \Delta},
\end{equation}
where
\begin{equation}
{\mathsf Q} = ({\mathsf R}^{t}{\mathsf R})^{-1}{\mathsf R}^{t}.
\end{equation} 
The resulting non-affinity $\chi$ from (\ref{chidef}) is
\begin{equation}
 \chi=\left({\bf \Delta}- {\mathsf R}({\mathsf R}^{t}{\mathsf R})^{-1}{\mathsf R}^{t}{\bf \Delta}\right)^{2} = {\bf \Delta}^{t}{\mathsf P}{\bf \Delta}
\label{chidef2}
\end{equation}
where
\begin{equation}
{\mathsf P} = {\mathsf I}-{\mathsf R}{\mathsf Q}
= {\mathsf I}-{\mathsf R}({\mathsf R}^{t}{\mathsf R})^{-1}{\mathsf
R}^{t}
\end{equation}
projects onto the space of ${\bf \Delta}$ that cannot be expressed as an affine deformation. 
Note that in arriving at (\ref{chidef2}) we have used the fact that ${\mathsf P}$ is symmetric, i.e.\ ${\mathsf P}^t = {\mathsf P}$ and 
\begin{eqnarray}
{\mathsf P}^2 & = & ({\mathsf I}- {\mathsf R}({\mathsf R}^{t}{\mathsf R})^{-1}{\mathsf R}^{t})^2 \nonumber \\
& = & {\mathsf I}^2 - 2\, {\mathsf R}({\mathsf R}^{t}{\mathsf R})^{-1}{\mathsf R}^{t} + \nonumber \\
& & {\mathsf R}({\mathsf R}^{t}{\mathsf R})^{-1}[({\mathsf R}^{t}{\mathsf R})({\mathsf R}^{t}{\mathsf R})^{-1}]{\mathsf R}^{t} \nonumber \\
& = & {\mathsf P}.
\end{eqnarray}
As usual this means that all eigenvalues of ${\mathsf P}$ are either
zero or one.

Having found explicit expressions for ${\bf e}$ and $\chi$ we now proceed to obtain their statistics at low temperature, where a harmonic approximation will be valid. Specifically we consider the canonical distribution of displacements ${\bf u}_i$ and momenta ${\bf p}_i$ at inverse temperature $\beta = 1/k_B T$:
\begin{equation}
P ({\bf p}_i, {\bf u}_i) = \frac{1}{Z}
\exp [-\beta H ({\bf p}_i, {\bf u}_i)]
\end{equation}
with the harmonic Hamiltonian
\begin{equation} 
H = \sum_{i} \frac{{\bf p}_i^2}{2 m_i} + \frac{K}{2} \sum_{(ij)} ({\bf u}_i - {\bf u}_j)^2,
\label{harm}
\end{equation}
where $m_i$ is the mass of particle $i$.
The sum in the second term in (\ref{harm}) runs over all bonds in a harmonic network with spring constants $K$. This is the Hamiltonian for the examples we consider in this paper. However, the general expressions that we derive apply directly also to generic quadratic Hamiltonians of the form
\begin{equation} 
H = \sum_{i} \frac{{\bf p}_i^2}{2 m_i}
+ \frac{1}{2\beta l^2} \sum_{i\alpha j\gamma} u_{i\alpha} {\mathcal D}_{i\alpha,j\gamma} u_{j\gamma}
\label{harm_gen}
\end{equation}
Here ${\mathcal D}_{i\alpha,j\gamma}$ is the dynamical matrix; we have
made this dimensionless by extracting a factor of $1/(\beta l^2)$.

Integrating out the momenta from the canonical distribution shows that the particle displacements have a Gaussian distribution. Their covariances can be expressed compactly in terms of the Fourier transform of the dynamical matrix~\cite{ashcroft, harmdyn} 
\begin{equation}
\Dmat_{\alpha\gamma}({\bf q}) =
\sum_{i} {\mathcal D}_{i\alpha,j\gamma}
e^{-i {\bf q} \cdot  ({\bf R}_{i}-{\bf R}_j)}
\end{equation}
where because of translational invariance the choice of reference particle $j$ is arbitrary. This matrix determines the variances of the Fourier components according to
\begin{equation}
\langle {\bf u}_{\bf q} {\bf u}_{-{\bf q'}}^t \rangle = \Dmat^{-1}({\bf q})
\,\vBZ\delta({\bf q}-{\bf q'})
\end{equation}
where the angled brackets indicate a thermal average. The covariances of the displacements are, accordingly,
\begin{equation}
\langle {\bf u}_i {\bf u}_j^t \rangle = 
l^2\int \frac{d{\bf q}}{\vBZ}\,\Dmat^{-1}({\bf q})\,
e^{i{\bf q}\cdot ({\bf R}_i -{\bf R}_j)}\ .
\end{equation}
For the particle displacements in our coarse-graining volume $\Omega$ of interest, ${\bf \Delta}_i = {\bf u}_i-{\bf u}_0$, we thus find also a Gaussian distribution with covariance matrix $\langle \Delta_{i\alpha}\Delta_{j\gamma}\rangle$ given by
\begin{eqnarray} 
C_{i\alpha,j\gamma}&=&l^2\int \frac{d{\bf q}}{\vBZ}\,
\Dmat^{-1}_{\alpha\gamma}({\bf q})(e^{i{\bf q}\cdot{\bf R}_{j}}-e^{i{\bf q}\cdot{\bf R}_{0}})\times \nonumber \\
& &(e^{-i{\bf q}\cdot{\bf R}_{i}}-e^{-i{\bf q}\cdot{\bf R}_{0}})
\label{dyn}
\end{eqnarray}
%
%
%
Note that the matrix ${\mathsf C}$ defined in this way has the symmetry of the lattice. The thermal average of any observable  $A(\bf {\Delta})$ is then given by, 
\begin{equation}
\langle A \rangle = \frac{1}{Z_\Omega} \int \prod_{i \alpha} d\Delta_{i \alpha} A({\bf \Delta}) \exp\left(-\frac{1}{2} {\bf \Delta}^{t}{\mathsf C}^{-1}{\bf \Delta}\right)
\label{expec}
\end{equation}
with normalization constant
$Z_{\Omega}=(2\pi)^{Nd/2}\sqrt{{\rm det}\,{\mathsf  C}}$.
%
In the next section we use (\ref{expec}) to obtain the probability distribution functions for $\chi$ and ${\bf e}$.

\section{Single point probability distributions}
\label{sec:2}
In this section we derive the single point (local) joint probability
distribution, $P(\chi,{\bf e})$, for non-affinity $\chi$ and strains
${\bf e}$. As before we consider lattices at non-zero but low temperatures where a harmonic approximation to particle interactions remains valid. To obtain $P(\chi,{\bf e})$, we begin with
\begin{eqnarray}
\varPhi(k,\bkappa) & = & \int d\chi\, d{\bf e}\, P(\chi, {\bf e}) \exp(ik\chi+i\bkappa^{t} {\bf e}) \nonumber \\
& = & \langle e^{ik\chi+i \bkappa^{t}{\bf e}}\rangle
\label{joint1}
\end{eqnarray} 
which is the characteristic function for the joint probability distribution $P(\chi, {\bf e})$ as measured within $\Omega$.  
Substituting the general expressions from Section~\ref{sec:1}, $\chi = {\bf \Delta}^{t}{\mathsf P}{\bf \Delta}$ and ${\bf e} = {\mathsf Q}{\bf \Delta}$, into (\ref{joint1}) we obtain
\begin{eqnarray}
\varPhi(k,\bkappa)& = &\frac{1}{Z_\Omega} \int \prod_{i \alpha} d\Delta_{i \alpha} \exp\Big[ {-\frac{1}{2}{\bf \Delta}^{t}{\mathsf C}^{-1}{\bf \Delta}} +  \nonumber \\
& & {i k {\bf \Delta}^{t}{\mathsf P}{\bf \Delta}} + {i \bkappa^{t} {\mathsf Q} {\bf \Delta}}\Big].
\label{joint2}
\end{eqnarray}
Completing the squares in the argument of the exponential in (\ref{joint2}) and carrying out the resulting Gaussian integrals yields
\begin{eqnarray}
 \varPhi(k,\bkappa)&=&\exp\left(-\frac{1}{2} \bkappa^{t}{\mathsf Q}{\mathsf C} ({\mathsf I}-2ik{\mathsf P}{\mathsf C})^{-1}{\mathsf Q}^{t}\bkappa\right) \times \nonumber \\
 & & [{\rm det}({\mathsf I}-2ik{\mathsf P}{\mathsf C}{\mathsf P})]^{-1/2}
 \label{phi}
 \end{eqnarray}
Setting $\bkappa=0$ and $k=0$ gives the characteristic functions of
$\chi$ and ${\bf e}$, respectively, as
\begin{eqnarray}
\varPhi_{\chi}(k) &=& [{\rm det}({\mathsf I}-2ik{\mathsf P}{\mathsf
  C}{\mathsf P})]^{-1/2} 
\label{single_char_chi}
\\
\varPhi_{\bf e}(\bkappa) &=& \exp\left(-\frac{1}{2}
  \bkappa^{t}{\mathsf Q}{\mathsf C}{\mathsf Q}^{t}\bkappa\right)
\label{single_char_e}
\end{eqnarray}
Extracting these factors from the joint characteristic function shows that it can be written as
\begin{eqnarray}
\varPhi(k,\bkappa)&=& \varPhi_{\chi}(k) \varPhi_{\bf e}(\bkappa)
\label{decomp_general}
\\
& & \times
\exp\left(-ik \bkappa^{t}{\mathsf Q}{\mathsf C} ({\mathsf I}-2ik{\mathsf P}{\mathsf C})^{-1}{\mathsf P}{\mathsf C}{\mathsf Q}^{t}\bkappa\right)
\nonumber
\end{eqnarray}
The term in the second line expresses the fact that $\chi$ and
${\bf e}$ are generally coupled to each other, rather than varying
independently. A special case where this does not happen occurs when
${\mathsf P}$ and ${\mathsf C}$ commute. Then one can
write ${\mathsf P}{\mathsf C}{\mathsf Q}^{t}= {\mathsf C}{\mathsf
P}{\mathsf Q}^{t}$. But this vanishes because from the definitions
of ${\mathsf P}$ and ${\mathsf Q}$ one has ${\mathsf P}{\mathsf
Q}^t=0$. The coupling term in (\ref{decomp_general}) then becomes
unity and $\chi$ and ${\bf e}$ are uncorrelated. This is the situation we will encounter in the one-dimensional example below, when coarse-graining on the smallest lengthscale where $\Omega$ only contains the nearest neighbours of particle 0.

In the case where ${\mathsf P}$ and ${\mathsf C}$ 
have a non-zero commutator $[{\mathsf P},{\mathsf C}] = 
{\mathsf P}{\mathsf C} - {\mathsf C}{\mathsf P}$, one can
put the expansion for small $k$ of the coupling factor in
(\ref{decomp_general}) into a form that emphasizes the role of this
commutator. Specifically, by writing ${\mathsf P}{\mathsf C} = 
{\mathsf C}{\mathsf P} - [{\mathsf P},{\mathsf C}]$ and exploiting 
the property ${\mathsf P}{\mathsf Q}^t=0$ one finds
%
%
\begin{widetext} 
\begin{eqnarray}
\varPhi(k,\bkappa)&=& 
\varPhi_{\chi}(k)\varPhi_{\bf e}(\bkappa)
\exp\left(-i\bkappa^{t}{\mathsf Q} {\mathsf C}\left[
[{\mathsf P},{\mathsf C}]\,k
+2i\Big( \left[{\mathsf C}{\mathsf P},[{\mathsf P},{\mathsf C}]\right]
+[{\mathsf P},{\mathsf C}]^{2} \Big)\,k^2  
+ \ldots \right]{\mathsf Q}^{t}\bkappa\right)
%
\label{full-phi}
\label{taylor}
\end{eqnarray}
\end{widetext}
%
%
%
%
%
From the general form (\ref{decomp_general}) of the characteristic
function, or its expanded version
(\ref{taylor}), we can then obtain the desired joint probability
distribution $P(\chi,{\bf e})$ by inverse Fourier transform, either
analytically or numerically.

Before proceeding to apply the above general
results to two simple example systems, we comment briefly on the
marginal distributions of $\chi$ and ${\bf e}$ whose characteristic
functions are given in (\ref{single_char_chi},\ref{single_char_e})
above. From the second of these equations, the distribution $P({\bf
e})$ of the local strain is a zero mean Gaussian distribution with
covariance matrix ${\mathsf Q}{\mathsf C}{\mathsf Q}^{t}$. For the
local non-affinity $\chi$, if we call $\sigma_j$ the eigenvalues of
the matrix ${\mathsf P}{\mathsf C}{\mathsf P}$, then 
the characteristic function (\ref{single_char_chi}) has the explicit
form
\begin{eqnarray}
\varPhi_{\chi}(k) & = &
\frac{1}{\prod_{j}\sqrt{1-2ik\,\sigma_{j}}}
\label{phichi}
\end{eqnarray}
This shows that $\chi$ has a generalized chi-square distribution
$P(\chi)$: it is a sum of squares of Gaussian random variables, each
with zero mean and variance $\sigma_j$. Only the nonzero $\sigma_j$
contribute here, and there are $Nd-d^2$ of these. This follows from
the fact that ${\mathsf P}$ eliminates from the space of all relative
displacements in $\Omega$A the $d^2$-dimensional subspace of
affine displacements.

\subsection{The one dimensional harmonic chain}
\label{subsec:chain}

\begin{figure}[h]
\begin{center}
\includegraphics[width=8.5cm]{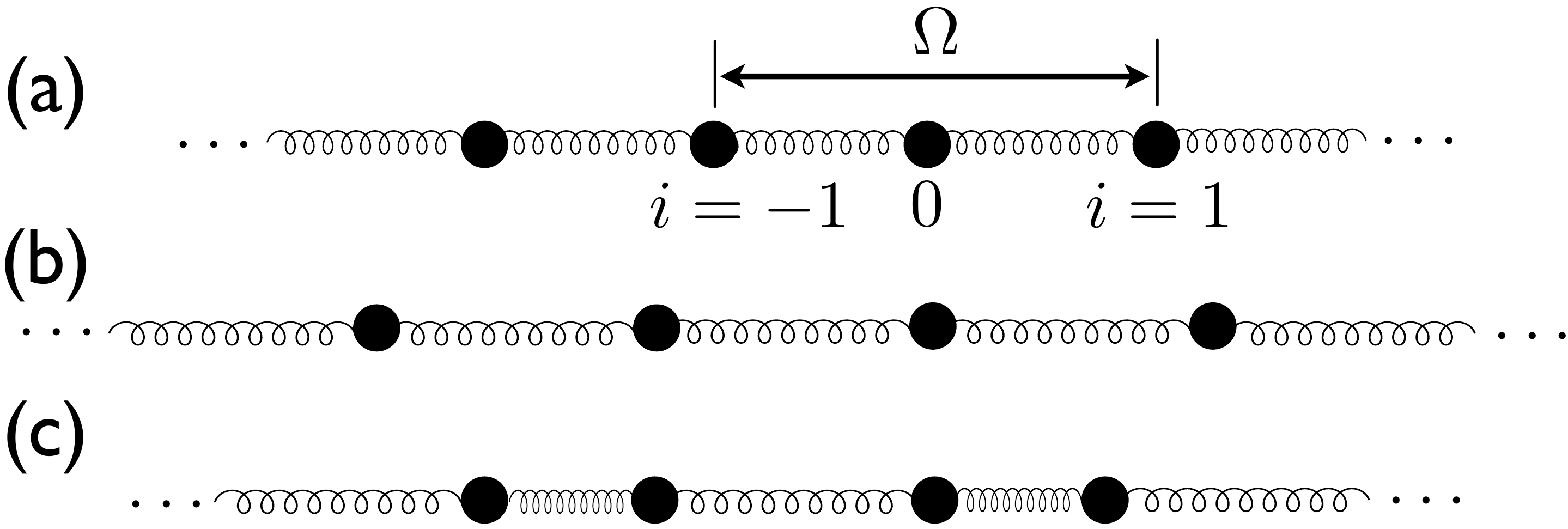}
\end{center}
\caption{(a) A portion of a one-dimensional harmonic chain showing the
  neighborhood $\Omega$ consisting of particle $0$ and its two nearest
  neighbors $i = \pm 1$. The affine (b) and the non-affine (c) modes
  are also shown.}
\label{1dlat}
\end{figure}

Consider a one-dimensional chain of particles of equal mass connected
by harmonic springs with spring constant $K$ and equilibrium length $l$
as shown in Fig.~\ref{1dlat}. We choose as the coarse-graining
neighborhood $\Omega$ a central particle $0$ at ${\bf R}_0 \equiv x_0
= 0$ and its two nearest neighbors at $x_{\pm 1}=\pm l$. Fluctuating
particle positions, ${\bf R}_i \equiv x_i$, produce displacements $u_i
= x_i - i\,l$ and the vector of relative displacements is
$\bm{\Delta}^t = (u_{1} - u_0\,,\, u_{-1} - u_{0})$. The matrices
defined in Section~\ref{sec:1}
can be easily evaluated for this system and are given by:
 \[ {\mathsf R} = \left( \begin{array}{c}
l \\
-l\end{array} \right), \]
\[{\mathsf Q}=({\mathsf R}^{t}{\mathsf R})^{-1}{\mathsf R}^{t} = \left( \begin{array}{cc}
\frac{1}{2l} & -\frac{1}{2l}\end{array} \right),
\] 
and 
\[{\mathsf P}={\mathsf I}-{\mathsf R}({\mathsf R}^{t}{\mathsf R})^{-1}{\mathsf R}^{t} = \left( \begin{array}{cc}
\frac{1}{2} & \frac{1}{2}\\ 
&\\
\frac{1}{2} & \frac{1}{2}\end{array} \right).
\]

The two eigenvectors of ${\mathsf P}$ corresponding to the eigenvalues
zero and one are $(l,-l)$ and $(l,l)$ respectively.  The mode
with the non-zero eigenvalue corresponds to the non-affine deformation
$\chi = (u_{1} + u_{-1} - 2 u_0)^2/2$, while the one corresponding to
the null space of ${\mathsf P}$ gives the
only affine mode ${\bf e} = \epsilon = (u_{1} - u_{-1})/2 l$ of the
lattice. The affine and the non-affine modes with respect to $\Omega$
are shown in Fig.~\ref{1dlat}(b) and (c) respectively.  The dynamical
``matrix'' in Fourier space is ${\Dmat}(q) = 2\, \beta Kl^2
\,[1-\cos(ql)]$; in the following we use energy units such that
$Kl^2=1$. The displacement covariance matrix (\ref{dyn}) then becomes
\begin{equation}
C_{ij} = \frac{l^2}{2 \beta}\frac{l}{2\pi}\int_{0}^{2\pi/l}\,dq \frac{(e^{iq\,{x}_{i}}-e^{iq\, x_{0}})
(e^{-iq\,x_{j}}-e^{-iq\, x_{0}})}{1-\cos(ql)}\\
\label{C_local_chain}
\end{equation}
which is simply $l^2 \beta^{-1}$ times the identity matrix and so 
\[{\mathsf P}{\mathsf C}{\mathsf P}^{t}= l^2 \beta^{-1}\left( \begin{array}{cc}
\frac{1}{2} & \frac{1}{2}\\
& \\
\frac{1}{2} & \frac{1}{2}\end{array} \right)
\]
with eigenvalues  $\sigma=l^2 \beta^{-1}$ and $0$, while ${\mathsf
  Q}{\mathsf C}{\mathsf
  Q}^{t}=\langle\epsilon^{2}\rangle=\beta^{-1}/2$.

The fact that $C_{ij}=l^2 \beta^{-1}\delta_{ij}$ as found above means
that the relative particle displacements are uncorrelated. This is
easy to see intuitively as the potential energy of the system is
$(K/2)\sum_{n=-\infty}^\infty (x_{n+1}-x_n-l)^2$. Relative
displacements $x_{n+1}-x_n-l$ of nearest neighbours therefore have
independent fluctuations, 
and the relative displacements $u_1-u_0=x_1-x_0-l$ and $u_{-1}-u_0 =
-(x_0-x_{-1}-l)$ in our coarse-graining neighborhood $\Omega$ are
exactly of this form. If we were to enlarge $\Omega$, say to include
next-nearest neighbours, then this would no longer hold as
e.g.\ $u_2-u_0=x_2-2l-x_0=(x_2-x_1-l)+(x_1-x_0-l)$ is correlated with
$u_1-u_0$.

Carrying out the matrix manipulations in (\ref{phi}) after
specializing to the $d=1$ case, we find that the $k$-dependence of the
first factor cancels out since $[{\mathsf P},{\mathsf C}] = 0$, in
line with the general discussion after (\ref{full-phi}). This yields
the characteristic function for the joint probability
distribution as the product of the individual characteristic
functions:
\begin{eqnarray}
 \varPhi(k,\kappa)&=&\left[\frac{1}{\sqrt{1-2ik\,\sigma}}\right]\left[\exp\left(-\frac{1}{2}\langle\epsilon^{2}\rangle\kappa^{2}\right)\right] \nonumber \\
&=&\varPhi_{\chi}(k)\varPhi_{\epsilon}(\kappa)
\end{eqnarray}
The joint probability distribution is then obtained by inverse Fourier transforming the characteristic function:
\begin{eqnarray}
P(\chi,\epsilon)&=&\left[\frac{1}{\sqrt{2\pi\,\sigma}}\chi^{-1/2}\exp\left(-\frac{\chi}{2\,\sigma}\right)\right] \times \nonumber \\
& &\left[\frac{1}{\sqrt{2\pi\langle\epsilon^{2}\rangle}}\exp\left(-\frac{\epsilon^{2}}{2\langle\epsilon^{2}\rangle}\right)\right] \nonumber \\
&=&P(\chi)P(\epsilon)
\end{eqnarray}
This has a simple form, namely, a product of the chi-square
distribution of a single Gaussian random variable and a Gaussian. We
can obtain immediately, for example, the $n$-th moments of $\chi$,
$\langle \chi^{n}\rangle =
(2\sigma)^{n}\Gamma(n+\frac{1}{2})/\Gamma(\frac{1}{2})$, which are all
finite. In Section\,~\ref{sec:4} we show that one can define an
external field $h_{\chi}$ that couples to $\chi$ and, for a specific value, can cause all the moments to diverge so that $P(\chi)$ crosses over to a distribution with a power-law tail.

\subsection{The two dimensional harmonic triangular net}

\begin{figure}[h]
\begin{center}
\includegraphics[width=5.0cm]{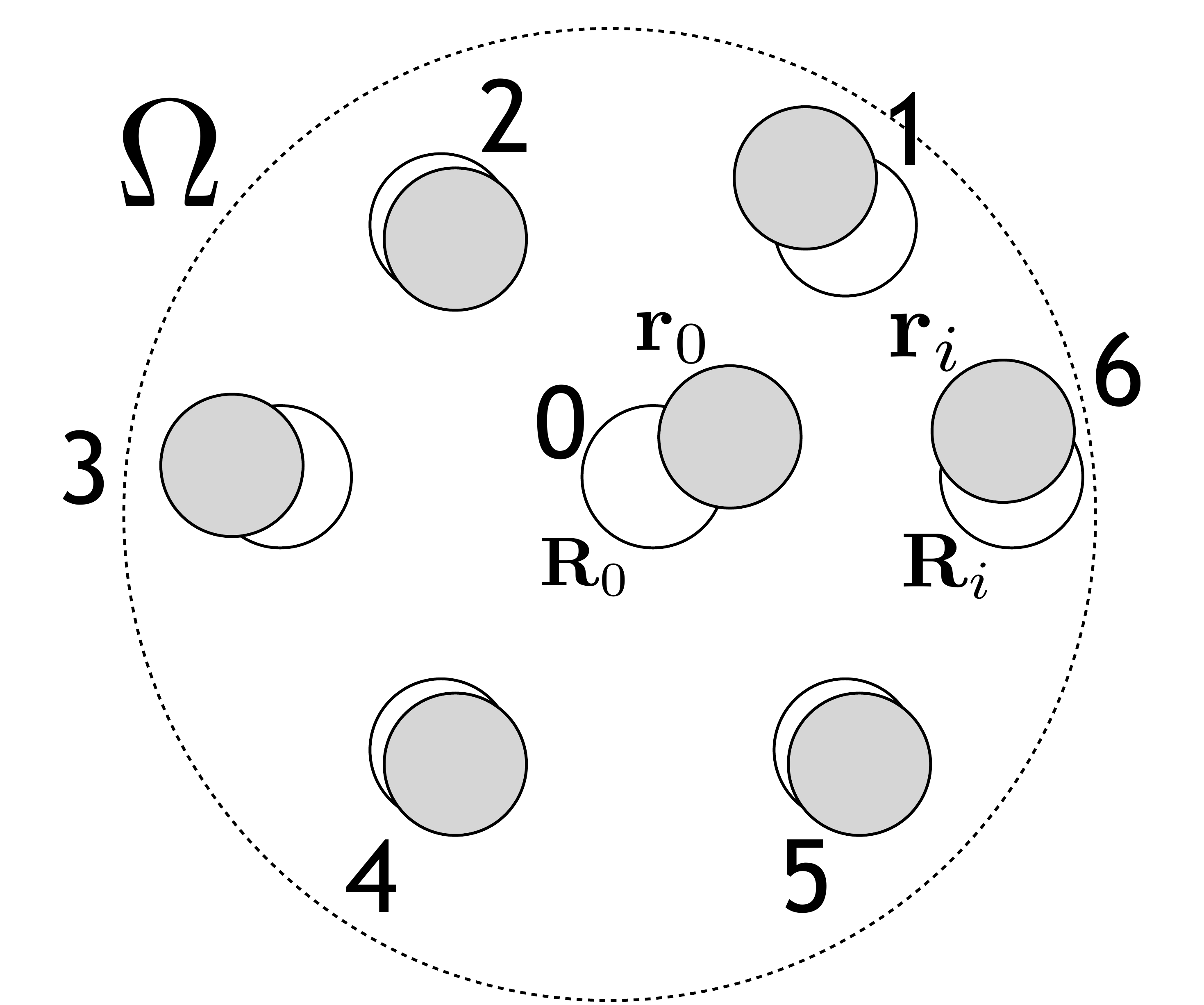}
\end{center}
\caption{A typical neighborhood $\Omega$ around a central particle
  $0$ in a triangular lattice, containing the six 
 nearest neighbor particles $i=1,\ldots,6$.
. The reference positions ${\bf R}_0$ and ${\bf R}_i$ are shown by
open circles while the instantaneous positions ${\bf r}_0$ and ${\bf
  r}_i$ are indicated by filled gray circles.}
\label{trilat}
\end{figure}

The joint probability distribution of local coarse-grained strain
${\bf e}$ and non-affinity $\chi$ for a two-dimensional triangular
lattice can be obtained in a similar manner. We choose again
a nearest neighbor (hexagonal) coarse-graining neighborhood $\Omega$ as shown in
Fig.~\ref{trilat}. To simplify the notation we also assume, without
loss of generality, ${\bf R}_0 = {\bf 0}$ in what follows, and take
the lattice constant as our length unit so that $l=1$.
Following the lines of the 1-d calculation, we begin by obtaining the matrices ${\mathsf R}$ and ${\mathsf P}$.

\subsubsection{${\mathsf R}$ and ${\mathsf P}$} The matrix ${\mathsf
  R}$ is a $12\times4$ matrix encoding the position vectors of the $6$
neighbors of 
the particle at the origin (see Fig.~\ref{trilat}). Explicitly one has
\begin{equation}
 {\mathsf R} = \left( \begin{array}{cccc}
R_{11} & R_{12} & 0 & 0 \\
0 & 0 & R_{11} & R_{12} \\
\vdots & \vdots & \vdots & \vdots \\
R_{61} & R_{62} & 0 & 0 \\
0 & 0 & R_{61} & R_{62}
\end{array} \right)
\end{equation} 
Here the index $\alpha = 1,2$ indicates the $x$ and $y$ components of
the lattice positions ${\bf R}_{i}$ respectively. 

To find the projection matrix ${\mathsf P}={\mathsf I}- {\mathsf
  R}({\mathsf R}^{t}{\mathsf R})^{-1}{\mathsf R}^{t}$, we substitute
the above form of ${\mathsf R}$ into the matrix ${\mathsf R}({\mathsf
  R}^{t}{\mathsf R})^{-1}{\mathsf R}^{t}$. One finds that this
consists of $6 \times 6$ blocks, each of which is a $2\times2$ diagonal
matrix of the form
\[
\frac{1}{3} (R_{i1}R_{j1} + R_{i2}R_{j2}) \left( 
 \begin{array}{cc}
1 & 0\\
0 & 1 \end{array}
\right)
= \frac{{\bf R}_i\cdot {\bf R}_j}{3}  \left( 
 \begin{array}{cc}
1 & 0\\
0 & 1 \end{array}
\right)
\]
The resulting ${\mathsf P}$ has $4$ zero eigenvalues corresponding to
the affine transformations. The $8$ unit eigenvalues correspond to
nonaffine distortions within $\Omega$. To identify a convenient basis
for the non-affine (8-dimensional) eigenspace, we choose below the
eigenvectors of ${\mathsf P}{\mathsf C}{\mathsf
P}$ with non-zero eigenvalues. Similarly, a physically meaningful
basis for the affine 
(4-dimensional) eigenspace is formed by the non-zero eigenvectors of
$({\mathsf I} - {\mathsf P}){\mathsf C} ({\mathsf I} - {\mathsf P})$.

\subsubsection{${\mathsf C}$ and ${\mathsf P}{\mathsf C}{\mathsf P}$}

In order to obtain the statistics of $\chi=\bm{\Delta}^t{\mathsf
  P}\bm{\Delta}$ and ${\bf e}={\mathsf Q}\bm{\Delta}$ we need to
calculate, as before, the eigenvectors and eigenvalues of the matrices ${\mathsf
  P}{\mathsf C}{\mathsf P}$ and ${\mathsf Q}{\mathsf C}{\mathsf Q}^t$.
As discussed above, these are the matrices determining the
characteristic functions 
(\ref{single_char_chi},\ref{single_char_e}) and hence the marginal
distributions.
We thus require the displacement correlation matrix
${\mathsf C}$, which in turn is calculated from the
Fourier-transformed dynamical matrix $\Dmat({\bf q})$.
For the Hamiltonian (\ref{harm}) of a regular harmonic triangular net
of particles with spring constant $K$ and lattice constant $l$ this is
\begin{widetext}
\begin{equation}
\Dmat({\bf q}) = \beta \,
\left( \begin{array}{cc}
3-2\,\cos(q_{x})-\cos(\frac{1}{2}q_{x})\cos(\frac{\surd3}{2}q_{y}) & 
\sqrt{3}\sin(\frac{1}{2}q_{x})\sin(\frac{\surd3}{2}q_{y})\\
\sqrt{3}\sin(\frac{1}{2}q_{x})\sin(\frac{\surd3}{2}q_{y}) & 
3[1-\cos(\frac{1}{2}q_{x})\cos(\frac{\surd3}{2}q_{y})]\\
\end{array}
\right) 
\end{equation}
\end{widetext}
Here we have again chosen energy units such that $Kl^2=1$. We also use
the more intuitive 
notation $q_x\equiv q_1$ and $q_y\equiv q_2$ for the wavevector components.

The elements of the real symmetric matrix ${\mathsf C}$ are obtained
by evaluating the integral (\ref{dyn}) over the Brillouin zone of the triangular
lattice. It can be shown, by utilising lattice symmetries, that the
integral can be transformed to one over a rectangular region:
\begin{eqnarray}
C_{i\alpha,j\gamma}  & = & \frac{1}{2 v_{BZ}}\int_{0}^{4\pi}dq_{x}\int_{0}^{\frac{4\pi}{\sqrt{3}}}dq_{y}\, \Dmat_{\alpha\gamma}^{-1}(q_x,q_y) \times \nonumber \\
& & (e^{i {\bf q \cdot R}_{i}}-e^{i {\bf q \cdot R}_{0}}) (e^{-i {\bf q \cdot R}_{j}}-e^{-i {\bf q \cdot R}_{0}})
\end{eqnarray}
We compute both the real and the imaginary parts of this
two-dimensional integral numerically, using $256$-point Gauss-Legendre
quadrature. The imaginary parts of the elements of ${\mathsf C}$
vanish and provide an estimate for the accuracy of our numerics. The
normalizing volume of the unit cell in the reciprocal lattice is $v_{BZ}
= \frac{8\pi^{2}}{\sqrt{3}}$.  ${\mathsf C}$ is a $6\times6$ block
matrix, with each block of size $2\times 2$ as before. 
Of course not all these 36 blocks need to be calculated independently,
because of the overall symmetry ${\mathsf C}={\mathsf C}^t$. Using
also the additional symmetry relations (see Fig.~\ref{trilat}) 
\begin{eqnarray}
C_{1 \alpha 1 \gamma} & = & C_{4 \alpha 4 \gamma},
C_{2 \alpha 2 \gamma}  =  C_{5 \alpha 5 \gamma},
C_{3 \alpha 3 \gamma}  =  C_{6 \alpha 6 \gamma} \nonumber \\
C_{1 \alpha 2 \gamma} & = & C_{4 \alpha 5 \gamma}, 
C_{2 \alpha 3 \gamma}  =  C_{5 \alpha 6 \gamma},
C_{3 \alpha 4 \gamma}  =  C_{6 \alpha 1 \gamma} \nonumber \\
C_{1 \alpha 3 \gamma} & = & C_{4 \alpha 6 \gamma}, 
C_{2 \alpha 4 \gamma}  =  C_{5 \alpha 1 \gamma}, 
C_{3 \alpha 5 \gamma}  =  C_{6 \alpha 2 \gamma}
\end{eqnarray}
one finds that only 12 blocks of ${\mathsf C}$ are distinct.

\begin{figure}[ht]
\begin{center}
\includegraphics[width=8.7cm]{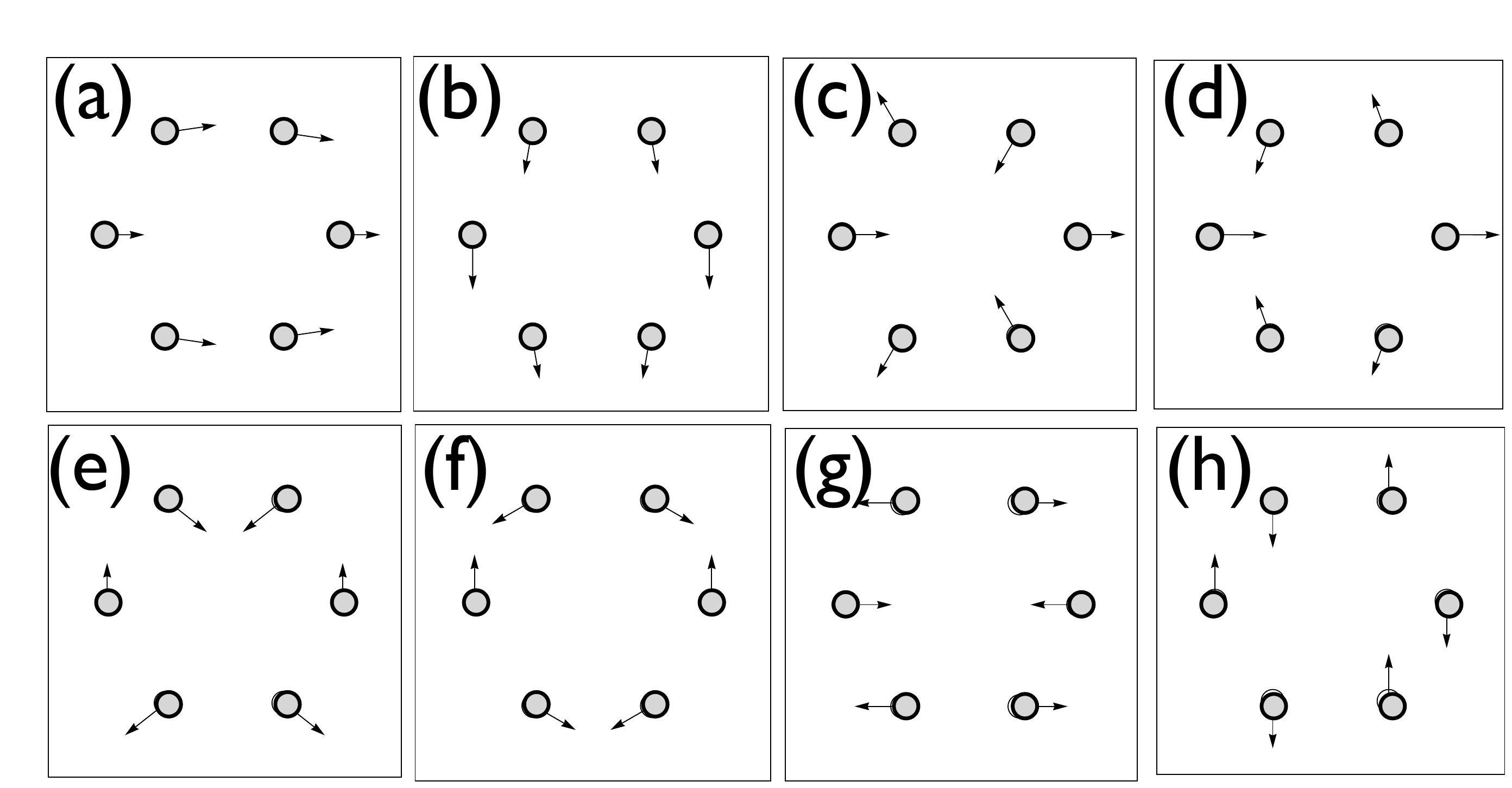}
\end{center}
\caption{Configurations showing the non-zero eigenvectors of ${\mathsf
    P}{\mathsf C}{\mathsf P}$, which represent non-affine
  displacements, corresponding to eigenvalues in descending order: (a,b) $\sigma_1 = \sigma_2$; (c) $\sigma_3$; (d,e) $\sigma_4 = \sigma_5$; (f,g,h) $\sigma_6=\sigma_7=\sigma_8$.}
\label{Fig.7}
\end{figure}

With ${\mathsf P}$ and ${\mathsf C}$ in hand one can construct and
diagonalize ${\mathsf P}{\mathsf C}{\mathsf P}$. This has $12$
eigenvalues $\sigma_j$ ($j=1,\ldots,12$), four of which are zero. The
$8$ non-zero eigenvalues, which correspond to the nonaffine
distortions within $\Omega$ shown in Fig.~\ref{Fig.7}, are
\begin{eqnarray}
\beta \sigma_1 & = & 2.454 = \beta \sigma_2\nonumber \\
\beta \sigma_3 & = & 0.482 \nonumber \\
\beta \sigma_4 & = & 0.312 = \beta \sigma_5 \nonumber \\
\beta \sigma_6 & = & 0.283 = \beta \sigma_7 = \beta \sigma_8 
\end{eqnarray}

\begin{figure}[ht]
\begin{center}
\includegraphics[width=8.7cm]{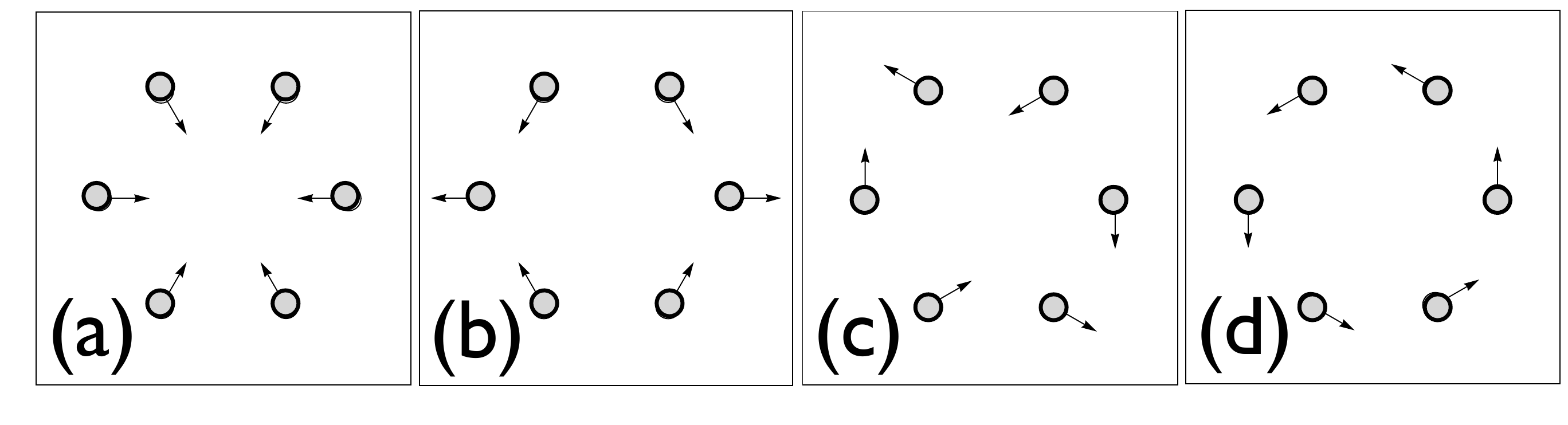}
\end{center}
\caption{Particle displacements for the non-zero
  eigenvectors of the matrix $({\mathsf I}-{\mathsf P}){\mathsf
    C}({\mathsf I}-{\mathsf P})$, showing affine eigendisplacements (a)
  dilation (b) uniaxial strain, (c) shear and (d) rotation. The
  reference lattice positions are shown by filled circles. The central atom has been deleted for clarity.}
\label{Fig.7a}
\end{figure}

The structure of the $4$-dimensional null space of ${\mathsf
  P}{\mathsf C}{\mathsf P}$ can be understood by looking at the
non-zero eigenvectors of the matrix obtained by the complementary
projection, viz.\ $({\mathsf I}- {\mathsf P}){\mathsf C}({\mathsf I} -
{\mathsf P})$. These eigenvectors correspond to the affine
eigendisplacements shown in Fig.~\ref{Fig.7a}.
%

\begin{figure}[h]
\begin{center}
\includegraphics[width=9.0cm]{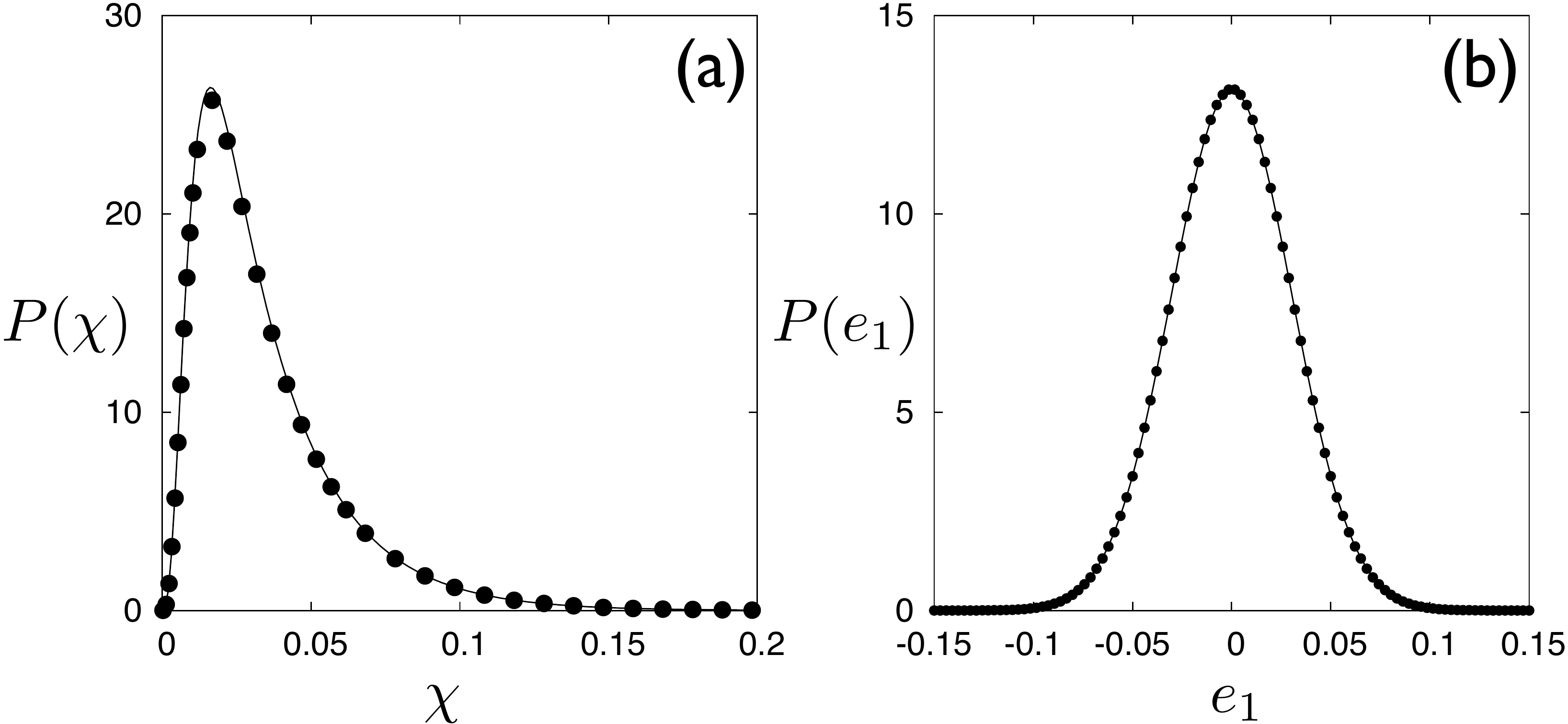}
\end{center}
\caption{(a) $P(\chi)$ from the exact calculation (line) compared with
  that obtained from molecular dynamics simulations (points) of a
  $400\times400$ site harmonic lattice with unit particle masses at reduced
  inverse temperature $\beta = 200$. The system was allowed to
  equilibrate for $2 \times 10^5$  MD steps with a timestep of
  $10^{-3}$, after which configurations were collected for $7 \times
  10^5$ MD steps. (b) Plot of $P(e_1)$ for the same system as in (a). 
}
\label{Fig.2}
\end{figure}

The independently fluctuating ``directions'' of the local
deformation tensor ${\bf e}$ can be worked out from the
covariance matrix ${\mathsf Q}{\mathsf C}{\mathsf Q}^t$ of the
Gaussian distribution of ${\bf e}$ (see (\ref{single_char_e}).
The projections of ${\bf e}$ onto these directions then have familiar
forms and map (see below) to the affine eigendistortions 
in Fig.~\ref{Fig.7a}. Explicitly we find for these projections:
(a) volume change (dilation), $e_{v} = e_1+e_4 =
D_{11}+ D_{22}$, (b) uniaxial strain, $\ed = e_1 - e_4 = D_{11}-
D_{22}$, (c) shear strain, $e_{s} = e_2 + e_3 = D_{12}+ D_{21}$, and 
(d) local rotation, $\omega = e_2 - e_3 = D_{12}-
D_{21}$. 
The associated eigenvalues give the relevant compliances for
our coarse-graining volume $\Omega$:
\begin{eqnarray}
\beta \langle e_v^2 \rangle & = & 0.261 \nonumber \\
\beta \langle \ed^2 \rangle & = & 0.481 = \beta \langle e_s^2 \rangle \nonumber \\
\beta \langle \omega^2 \rangle & = & 0.699.
\end{eqnarray} 
The statistics of the local deformation tensor $P({\bf e})$ therefore
consists of independent Gaussian fluctuations of these 4 deformation
modes, as illustrated in Fig.~\ref{Fig.2}(b).


%

To finish our discussion of the affine displacements we comment
briefly on the relation between the eigenvectors of $({\mathsf I}-
{\mathsf P}){\mathsf C}({\mathsf I} - {\mathsf P})$, which give the
independently fluctuating displacement patterns in the affine subspace
(``affine eigendisplacements''), and the eigenvectors of ${\mathsf
Q}{\mathsf C}{\mathsf Q}^t$, which are the independently fluctuating
components of the local deformation tensor (``eigendistortions''). The
two matrices are related via
\begin{equation}
({\mathsf I}- {\mathsf P}){\mathsf C}({\mathsf I} -
{\mathsf P}) = {\mathsf R}({\mathsf Q}{\mathsf C}{\mathsf Q}^t) {\mathsf R}^t
\label{compl}
\end{equation}
In the discussion above we treated their eigenvectors on the same
footing, and indeed each eigendistortion ${\bf \hat{e}}$ as an
eigenvector of ${\mathsf Q}{\mathsf C}{\mathsf Q}^t$ is related to an
affine eigendisplacement given by $\hat{\bf \Delta}={\mathsf R}{\bf
\hat{e}}$. This works, i.e.\ ${\mathsf R}{\bf \hat{e}}$ is indeed an
eigenvector of (\ref{compl}), because ${\mathsf R}^t{\mathsf R}$ is a
multiple of the identity matrix in the two-dimensional triangular
net. This simplification will hold in all lattices with sufficiently
high symmetry. Indeed, one can check that ${\mathsf R}^t{\mathsf R}$
has a $d\times d$ block structure where the off-diagonal blocks are
zero and the diagonal blocks are all equal to $\sum_i {\bf R}_i {\bf
R}_i^t$. This diagonal block is a matrix that commutes with the
entire symmetry group of the lattice so by Schur's lemma~\cite{tinkham} will be
proportional to the identity matrix, unless the symmetry group of the
lattice is too small.

Next we look at the statistics of the non-affinity parameter
$\chi$. The characteristic function (\ref{single_char_chi}) can be
written out explicitly as in (\ref{phichi})
where the $\sigma_{j}$s are the eigenvalues of  ${\mathsf P}{\mathsf C}{\mathsf P}$.
As we saw above, this matrix has eight
non-zero eigenvalues $\sigma_1,\ldots,\sigma_8$ and four zero
eigenvalues $\sigma_9=\ldots=\sigma_{12}=0$ that do not contribute to
(\ref{phichi}) as they correspond to purely affine distortions within $\Omega$. The eigenvectors associated with the $8$ non-affine displacements are shown in Fig.\ref{Fig.7}. 
Thus, as discussed above in general terms,
$P(\chi)$ is the distribution of the 
sum of the squares of  $Nd-d^{2}=8$ uncorrelated Gaussian random
variables, with the variances of these Gaussians given by the
eigenvalues $\sigma_{1}$, \ldots, $\sigma_8$. A numerical Fourier
transform of (\ref{phichi}) then gives $P(\chi)$. The result 
is shown in Fig.~\ref{Fig.2}(a), where we also 
compare with data from
molecular dynamics simulations; the agreement is evidently very good.

The first and the second moments of $\chi$ may be obtained from
successive derivatives of $\varPhi _{\chi}(k)$ with respect to its
argument so that $\langle\chi\rangle= i^{-1} \left(d
  \varPhi_{\chi}(k)/dk\right)_{k=0}=\sum_{j=1}^{8}\sigma_{j} = 6.865
\beta^{-1}$  and
\begin{eqnarray}
\langle\chi^{2}\rangle - \langle \chi \rangle^2 &=&-\left(\frac{d^{2}}{dk^{2}}\varPhi_{\chi}(k)\right)_{k=0} - \langle \chi \rangle^2 \nonumber \\
&=&  2 \sum_{j=1}^{8}\,\sigma_{j}^{2} \nonumber \\
&=&25.426\,\beta^{-2}.
\end{eqnarray}
The values for the corresponding  quantities obtained from our MD
simulations of $160,000$ particles and $7,000$ independent
configurations are $(6.869 \pm 0.005)\beta^{-1}$ and $(25.51 \pm
0.1)\beta^{-2}$. These are in excellent agreement with our
theoretical results; the agreement in all other quantities shown below
is of similar quality. Finally, the $n^{th}$ cumulant of $\chi$ is given by $(1/2)(n-1)! \sum_{j}(2\sigma_{j})^n$.

 \begin{figure}[h]
\begin{center}
\includegraphics[width=8.5cm]{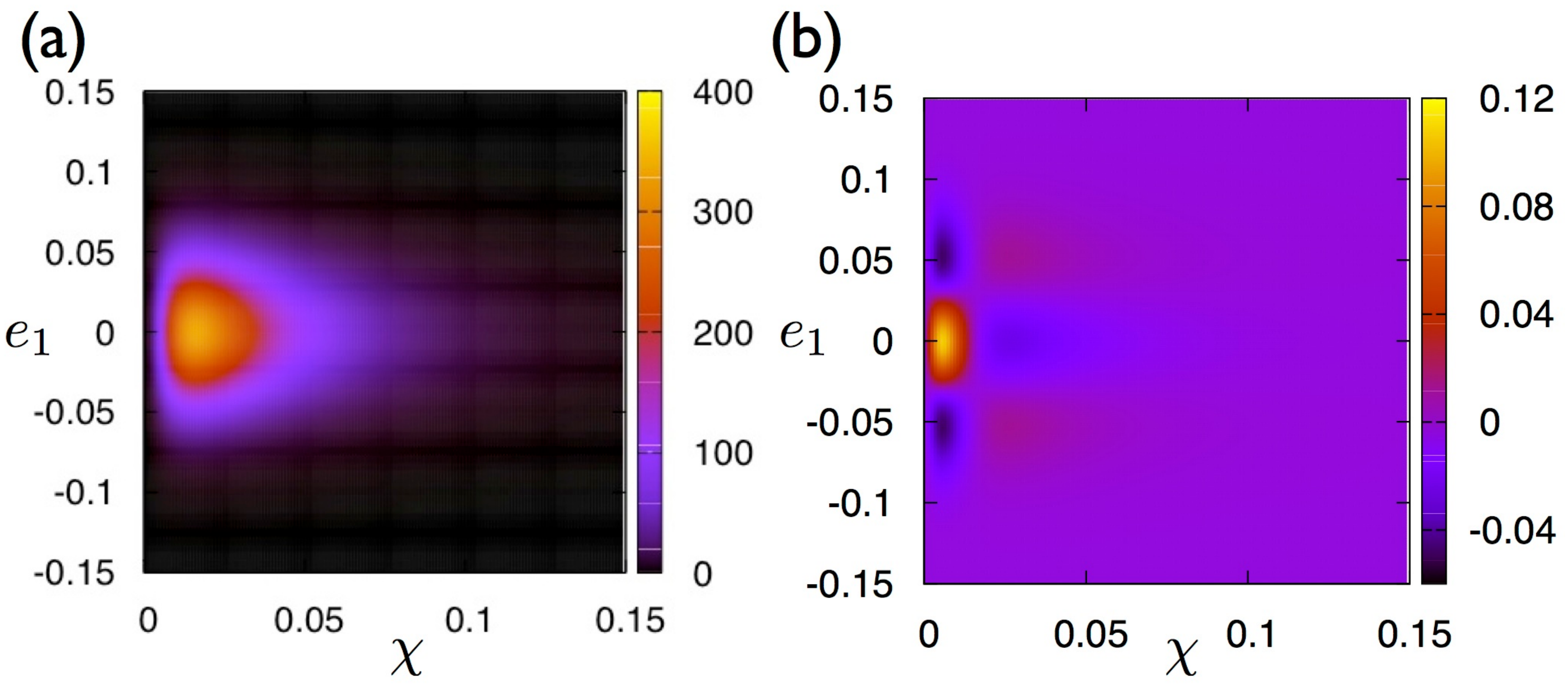}
\end{center}
\caption{(color-online) (a) Plot of $P(\chi, e_{1}) \approx
  P(\chi)P(e_1)$ as obtained from our calculations neglecting cross
  correlations and (b) the correction term $P(\chi, e_1) -
  P(\chi)P(e_1)$, both at $\beta = 200$. Note that the correction term is nonzero though small. The results from our simulations for $P(\chi, e_1)$ are indistinguishable from (a). The plots for other components of $\bf e$ are similar.}
\label{Fig.3}
\end{figure}
Having looked at the distributions of local strain ${\bf e}$ and
non-affinity $\chi$ separately, we finally ask about their correlations.
One can verify that, though small, the commutator $[{\mathsf
  P},{\mathsf C}]$ is non-vanishing, in contrast to the
one-dimensional harmonic chain case. Indeed, measuring matrix sizes by
the Euclidean norm $\lVert{\mathsf A} \rVert = \sqrt{{\rm Tr}\,
  {\mathsf A}{\mathsf A}^t}$,  we obtain for the chosen
nearest-neighbour coarse graining volume $\Omega$
\begin{align}
 \beta \lVert{\mathsf C} \rVert&=&3.875 \nonumber \\
\beta \lVert[{\mathsf P},{\mathsf C}]\rVert&=&0.124 \nonumber \\
\beta \lVert\left[{\mathsf C}{\mathsf P},[{\mathsf P},{\mathsf C}]\right]\rVert&=&3.591\times 10^{-2}
\end{align}
Neglecting the commutator to first approximation gives a joint
distribution of $\chi$ and ${\bf e}$ that factorizes into 
$P(\chi)$ and $P({\bf e})$ without any correlation. The result of this calculation for the joint probability $P(\chi,e_1)$ is shown in Fig.~\ref{Fig.3}(a). 
In actual fact, however,
correlations are present.
%
%
This means that the effective compliance of the solid depends on the
value of $\chi$ (and vice-versa). The effect is quantitatively rather
small for our example system, as is clear from
Fig.~\ref{Fig.3}(b) where we plot the first correction to the
factorized approximation. 
More simply, the
coupling between $\chi$ and ${\bf e}$ can be assessed by looking at
cross-correlations like
\begin{equation}
\langle\chi {\bf e}{\bf e}^t\rangle-\langle\chi\rangle\langle{\bf
  e}{\bf e}^t\rangle = 2\,{\mathsf Q}{\mathsf C}[{\mathsf P},{\mathsf
  C}]{\mathsf Q}^{t}.
\label{third_order}
\end{equation}
This result follows from the fact that the l.h.s.\ is a third-order
cumulant because $\langle {\bf e}\rangle=0$, and so is proportional to the
coefficient of the $O(k\bkappa^2)$ term in the
expansion (\ref{taylor}) of $\ln\varPhi(k,\bkappa)$. The commutator $[{\mathsf P},{\mathsf
  C}]$ only appears linearly here; higher orders
would be needed to express correlations involving higher
moments such as $\langle (\chi-\langle \chi\rangle)^n{\bf e}{\bf
e}^t) \rangle$. 
Using the numerically calculated $[{\mathsf P},{\mathsf C}]$ in
(\ref{third_order}) we obtain the following third-order
cross-correlation of $\chi$ with the eigendistortions: 
\begin{eqnarray}
\langle\chi e_v^2\rangle-\langle\chi\rangle\langle e_v^2 \rangle & = & \langle\chi \omega^2\rangle-\langle\chi\rangle\langle \omega^2 \rangle = 0  \label{cross}
\\
\langle\chi \ed^2\rangle-\langle\chi\rangle\langle \ed^2 \rangle & = &  \langle\chi e_s^2\rangle-\langle\chi\rangle\langle e_s^2 \rangle\,\, =  5.108 \times 10^{-3} \beta^{-2} \nonumber \\
\end{eqnarray}
One reads off that non-affinity is coupled with uniaxial
strain and shear strain, while local volume strain and rotational
distortions do not generate non-affinity. (This remains true also for
correlations involving higher orders of $\chi$.) The correlations
of $\chi$ with $\ed$ and $e_s$ are positive, so that a large affine
strain locally is generally accompanied by a large non-affinity
$\chi$. Conversely, a large value of $\chi$ makes the local region
more elastically compliant, i.e.\ typically leads to larger affine strains
${\bf e}$. 
More explicitly, the affine
displacements $({\mathsf I}-{\mathsf P})\bm{\Delta}$ conditional on
the non-affine ones ${\mathsf P}\bm{\Delta}$ can be written as a
linear function of ${\mathsf P}\bm{\Delta}$ plus Gaussian
fluctuations that do not depend on ${\mathsf P}\bm{\Delta}$. Thus one
can find the distribution of the affine displacements, and hence of
${\bf e}$, conditional on $\chi$ from the distribution of the first contribution across all non-affine displacements satisfying $({\mathsf P}\bm{\Delta})^2=\chi$. Because the first contribiution is linear in $\Delta$, one deduces that ${\bf e}$ is a sum of a random contribution proportional to $\chi^{1/2}$, and an independent Gaussian contribution.
From this it follows, for example, that
$\langle \chi^n{\bf e}{\bf e}^t \rangle =
\langle\chi^{n+1}\rangle {\bf M}_1 + \langle \chi^{n}\rangle {\bf M}_2$ where ${\bf M}_1$ and ${\bf M}_2$ are two $n$-independent matrices related to ${\mathsf C}$.

The strength of the coupling between $\chi$ and ${\bf e}$ 
will of course depend on the chosen
coarse-graining region $\Omega$, and in particular on its radius
$R_{\Omega}$; we return to this topic in Section~\ref{sec:5}. But
we believe that the positive correlation between the strengths of
local affine and non-affine deformations is 
not specific to the harmonic system considered here and should hold
generally for all solids. 
 
   

\section{Two point distributions and correlation functions}
\label{sec:3}
\begin{figure}[h]
\begin{center}
\includegraphics[width = 7.0cm]{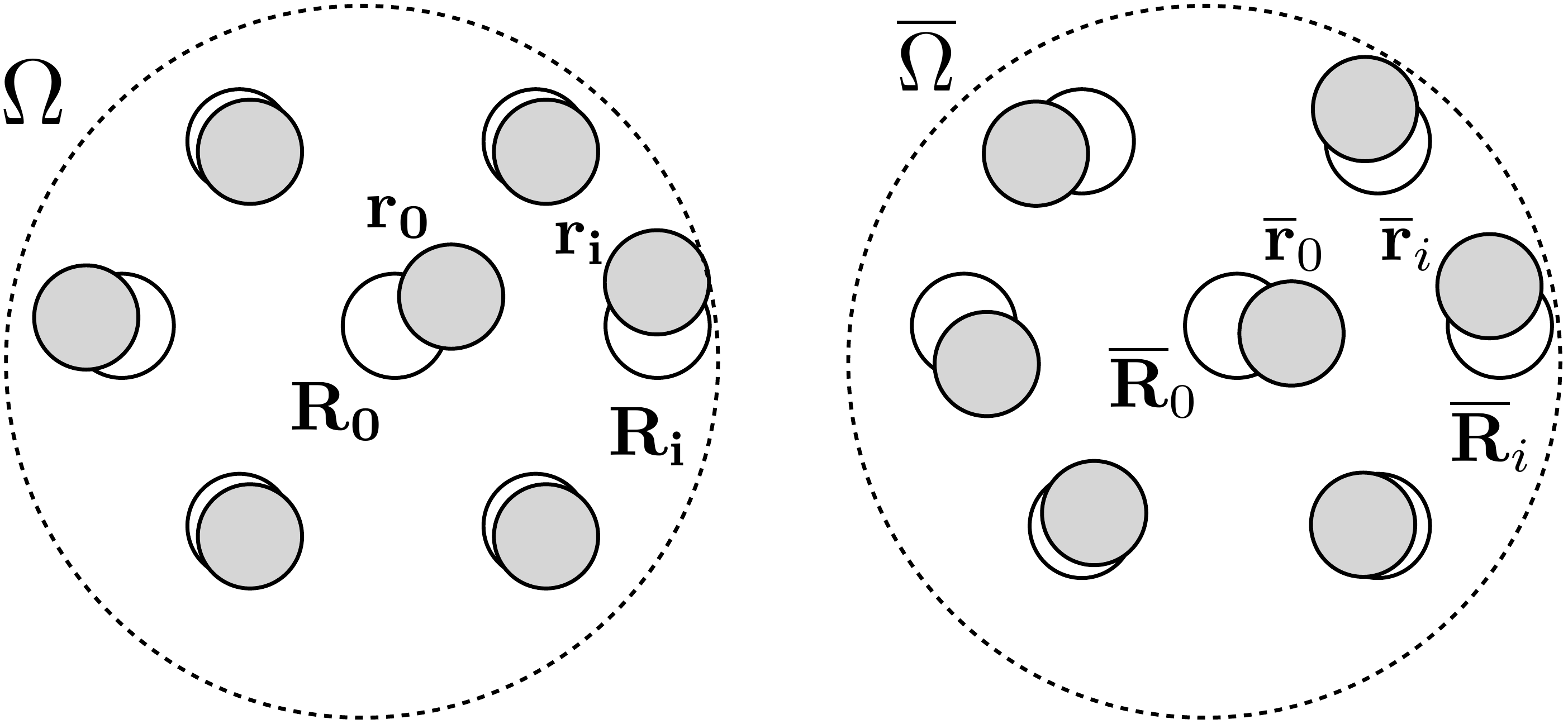}
\end{center}
\caption{Typical neighborhoods $\Omega$ and $\bar \Omega$ around particles $0$ and $\bar 0$ in the $d=2$ triangular lattice, illustrating the definitions used for obtaining the two-point correlation functions. The labels have the same meanings as in Fig.~\ref{trilat}.}
\label{trilat2}
\end{figure}

We now turn our attention to the spatial correlations of $\chi$ and
${\bf e}$. This requires us to consider simultaneously the
displacement differences 
in {\em two} neighborhoods $\Omega$ and $\bar \Omega$ centered on
lattice positions $\bR_0$ and $\barR_0$, respectively. The vector
${\bf \Delta}$ is defined as above, with an analogous definition for
$\bar{\bf \Delta}$. The
geometry and notation for the $d=2$ triangular lattice are given in
Fig.~\ref{trilat2}; the $d=1$ case is straightforward. The local
affine strain ${\bf e} = {\mathsf Q} {\bf \Delta}$ and non-affinity
$\chi = {\bf \Delta}^t {\mathsf P} {\bf \Delta}$ around $\bR_0$ are
then as before for $\Omega$ whereas for $\bar \Omega$ we have the corresponding quantities  $\bar {\bf e} = \bar{ {\mathsf Q}} \bar{ {\bf \Delta}}$ and $\bar \chi = \bar {{\bf \Delta}}^t \bar{{\mathsf P}}\bar{ {\bf \Delta}}$. Note that since the reference $\bar\Omega$ is just a translated copy of $\Omega$, one has $\bar{{\mathsf Q}}={\mathsf Q}$ and $\bar{{\mathsf P}} = {\mathsf P}$. 

To obtain the joint distribution of $\chi$, ${\bf e}$, $\bar\chi$ and
$\bar{\bf e}$ we need the joint Gaussian distribution of the
displacements ${\bf \Delta}$ and $\bar{\bf \Delta}$. These have covariances
\begin{eqnarray}
C_{i\alpha,j\gamma}&=&\langle \Delta_{i\alpha}\Delta_{j\gamma}\rangle,  \nonumber \\
{\bar {\bar {C}}}_{i\alpha,j\gamma}&=&\langle \bar{\Delta}_{i\alpha}\bar{\Delta}_{j\gamma}\rangle, \nonumber \\
\bar {C}_{i\alpha,j\gamma}&=&\langle \Delta_{i\alpha}\bar{\Delta}_{j\gamma}\rangle.
\end{eqnarray}
While the first two averages are identical, corresponding to two
different but equivalent lattice sites, the third quantity encodes the
displacement correlations between the two different sites. It may be obtained from an expression similar to~(\ref{dyn}),
\begin{eqnarray} 
\bar {C}_{i\alpha,j\gamma} & = & l^2\int \frac{d{\bf q}}{\vBZ} \,
\Dmat_{\alpha\gamma}^{-1}({\bf q})(e^{i {\bf q \cdot R}_{i}}-e^{i {\bf q \cdot R}_{0}}) \times \nonumber \\
& & (e^{-i {\bf q \cdot \bar{R}}_j}-e^{-i {\bf q \cdot \bar{R}}_{0}}).
\label{corr}
\end{eqnarray}
Note that $\bar {C}_{i\alpha,j\gamma}$ is not symmetric with respect
to interchanging $\bR_0$ and $\barR_0$, although the correlation
functions obtained from it below are, as they must be.

We could now proceed as for the local distribution $P(\chi,{\bf e})$
and derive the characteristic function
$\varPhi(k,\bkappa,\bar{k},\bar{\bkappa})$ of the joint distribution
$P(\chi,{\bf e},\bar{\chi},\bar{\bf e})$. This contains rather too
much information to present in a concise manner, however, so we
focus directly on the correlation functions. The simplest one of
these is the strain-strain correlator
\begin{equation}
\langle {\bf e} \bar {\bf e}^t\rangle = \langle {\mathsf Q} {\bf
  \Delta} \bar {\bf \Delta}^t  {\mathsf Q}^t \rangle =  {\mathsf Q}
\bar {\mathsf C} {\mathsf Q}^t
\end{equation}
In order to obtain space dependent correlation functions, this
quantity needs to be evaluated for all choices of
$\barR_0$ for fixed ${\bf R}_0$ (which may again be taken as the
origin).

Alternatively, one may obtain the correlation functions in {\bf
  q}-space. As a side effect, this avoids the Brillouin
zone integration in (\ref{corr}). To see this, write
\begin{eqnarray} 
\bar {C}_{i\alpha,j\gamma} & = & l^2 \int \frac{d{\bf q}}{\vBZ}\,
\Dmat_{\alpha\gamma}^{-1}({\bf q})(e^{i {\bf q} \cdot (\bR_i-\bR_0)}-1) \times \nonumber \\
& & (e^{-i {\bf q} \cdot (\barR_j-\barR_0)}-1) e^{i {\bf q} \cdot (\bR_0-\barR_0)}.
\label{corr2}
\end{eqnarray}
Now because the reference positions in $\Omega$ and $\bar\Omega$ are
just translated copies of each other, one has
$\barR_j-\barR_0=\bR_j-\bR_0$, so that the only $\barR_0$-dependence
resides in the last factor. Defining the Fourier transform of 
$\bar {C}_{i\alpha,j\gamma}$ via $\bar {C}_{i\alpha,j\gamma} =
l^2\vBZ^{-1} \int d{\bf q}\, \bar{C}_{i\alpha,j\gamma}({\bf q}) e^{i {\bf
    q} \cdot (\bR_0-\barR_0)}$ one thus reads off
\begin{eqnarray} 
\bar {C}_{i\alpha,j\gamma}(\bf q) & = & 
\Dmat_{\alpha\gamma}^{-1}({\bf q})(e^{i {\bf q} \cdot (\bR_i-\bR_0)}-1) \times \nonumber \\
& & (e^{-i {\bf q} \cdot (\barR_j-\barR_0)}-1).
\label{cor-q-space}
\end{eqnarray}
The Fourier transform of the strain correlation functions is then
simply
\begin{equation}
\langle {\bf e} \bar {\bf e}^t\rangle(\bf q) = 
{\mathsf Q} \bar {\mathsf C}(\bf q){\mathsf Q}^t
\label{strain_FT_general}
\end{equation}
and can be written down in closed form provided the dynamical matrix
$\Dmat(\bf q)$ for the lattice is known.


Next we consider the spatial correlation functions of the
non-affinity, $\langle \chi\barchi\rangle -
\langle\chi\rangle \langle\barchi\rangle$. 
It is convenient, at this stage, to define the vectors ${\bf Y} = {\mathsf P} {\bf \Delta}$ and $\bar {\bf Y} = {\mathsf P} \bar {\bf \Delta} $ 
with components $y_{j}$ and $\bar {y}_{j}$ where $j=1, \dots, Nd$. Thus $\chi  = \sum_{j=1}^{Nd}y_{j}^{2}$ and 
$\chi\barchi = \sum_{i,j=1}^{Nd}y_{i}^{2}\bar{y}_{j}^{2}$.
 Hence the correlation between $\chi$ and $\barchi$ is given by
\begin{eqnarray}
\langle \chi\barchi\rangle -  \langle\chi\rangle\langle\barchi\rangle
&=&\sum_{i,j=1}^{Nd}(\langle y_{i}^{2}\bar{y}_{j}^{2}\rangle  - 
\langle y_{i}^{2}\rangle\langle\bar{y}_{j}^{2}\rangle)
\nonumber
\\
& = &
2 \sum_{i,j=1}^{Nd}\langle y_{i} \bar{y}_{j}\rangle^2 
\ = \ 
2 \sum_{i,j=1}^{Nd} ({\mathsf P}\bar{\mathsf C}{\mathsf P})_{ij}^2
\nonumber
\\
& = &  
2\,{\rm Tr} ({\mathsf P}\bar{\mathsf C}{\mathsf P})
({\mathsf P}\bar{\mathsf C}{\mathsf P})^t
\end{eqnarray}
using Wick's theorem. If in line with our earlier notation
we use $\bar \sigma_j^2$ to denote the $Nd$ eigenvalues of the 
matrix $({\mathsf P}\bar{\mathsf C}{\mathsf P})
({\mathsf P}\bar{\mathsf C}{\mathsf P})^t$, then the last expression
can be simplified to
\begin{eqnarray}
\langle \chi\barchi\rangle -  \langle\chi\rangle^{2}&=&
 2\sum_{j=1}^{Nd}\bar {\sigma}_{j}^{2}
\label{chicor}
\end{eqnarray}
Note that if ${\mathsf P}\bar{\mathsf C}{\mathsf P}$ itself happens to be
symmetric, then the $\bar \sigma_j^2$ can be obtained as the squares of the
eigenvalues of this matrix. As for the
real-space strain correlator, one has to evaluate (\ref{chicor}) for
differt choices of $\barR_0$ to obtain spatial profiles of the
non-affinity correlator.
%
%

Finally one could ask about spatial cross-correlations like 
$\langle\chi \bar{\bf e}\bar{\bf e}^t\rangle-\langle\chi\rangle\langle\bar{\bf
  e}\bar{\bf e}^t\rangle$. We do not pursue this here: as we saw above, these
correlations are already rather weak (at least for coarse graining
across nearest neighbours) locally, i.e.\ when $\bR_0=\barR_0$.

\subsection{The one dimensional harmonic chain}

We now apply the above framework to the one-dimensional harmonic chain
introduced in Sec.~\ref{subsec:chain}. We choose as the first
reference location $x_0=0$ and as the second $\bar{x}_0=ml$. Coarse
graining will be across the nearest neighbours $x_{\pm 1}$ in $\Omega$
and $\bar{x}_{\pm 1}=(m\pm 1)l$ in $\bar\Omega$.
The matrix $\bar{C}_{jk}$ is then given by, 
\begin{equation}
\bar{C}_{jk}=l^{2}\int_{0}^{2\pi/l}\frac{dq}{2\pi/l}\, \frac{F(q)}{\Dmat(q)}
\end{equation}
where
\begin{eqnarray}
F(q)&=&(e^{iq\,{x}_{j}}-1)
(e^{-iq\,\bar{x}_{k}}-e^{-iq\,\bar{x}_{0}}) \\
&=&e^{-iq\bar{x}_{0}}\left(e^{iq(j-k)l}-e^{iq j l}-e^{-iq k
    l}+1\right)
\nonumber
\end{eqnarray}
There are now two possibilities. If $j = k$ then, bearing in mind that
$\Dmat(q) = 2\beta[1-\cos(ql)]$, one has $F(q)/\Dmat(q) =
\beta^{-1}e^{-iq\bar{x}_{0}}$ so that $\bar{C}_{jk}=0$ except when
$\bar{x}_{0}=0$. Otherwise, if $j = -k$, then
$F(q)/D(q)=-\beta^{-1}e^{iq(j l-\bar{x}_{0})}$ so that now
$\bar{C}_{jk}=0$ unless $\bar{x}_{0}=j l$. For all other cases,
$\bar{C}_{jk}$ vanishes identically. 
Summarizing, $\bar{\mathsf C}$ equals ${\mathsf C}$ from (\ref{C_local_chain}) when
$\bar{x}_0=0=x_0$; for $\bar{x}_{0}= l$ it is given by
\begin{equation}
\bar{{\mathsf C}}=l^{2}\beta^{-1} \left( \begin{array}{cc}
0 & -1\\
0 & 0\end{array} \right)
\label{Cbar_chain}
\end{equation}
while for $\bar{x}_0=-l$ one obtains the transpose. For all larger
distances $|\bar{x}_0|\geq 2l$, $\bar{\mathsf C}=0$, indicating that
$\chi$ and $\epsilon$ are uncorrelated beyond nearest neighbours.
The intuition here is as discussed after (\ref{C_local_chain}), namely
that the relative particle displacements of all nearest neighbour
pairs fluctuate independently from each other. Accordingly,
the single nonzero entry in (\ref{Cbar_chain}) comes from the correlation of the displacements
$u_1=x_1-x_0$ and $\bar{u}_{-1}=\bar{x}_{-1}-\bar{x}_0=x_0-x_1$, and
so is simply the negative variance of $u_1$.

To obtain the correlation functions we need the matrices ${\mathsf
  P}\bar{{\mathsf C}}{\mathsf P}^{t}$ and ${\mathsf
  Q}\bar{{\mathsf C}}{\mathsf Q}^{t}$, where we can focus
directly on the only non-zero correlations at $\bar{x}_{0}=\pm
l$. Even though $\bar{\mathsf C}$ is then not symmetric, 
${\mathsf P}\bar{{\mathsf C}}{\mathsf P}^{t}$ is, and has
eigenvalues $\bar{\sigma} = \frac{1}{2}l^{2}\beta^{-1}$ and $0$. On
the other hand, the scalar ${\mathsf Q}\bar{{\mathsf C}}{\mathsf
  Q}^{t}$ equals $\frac{1}{4}\beta^{-1}$. Using (\ref{chicor}), the
correlation functions, $\langle\chi(0)\chi(ml)\rangle - \langle \chi
\rangle^2$ and $\langle\epsilon(0)\epsilon(ml)\rangle$, can then be written as follows:
\begin{align*}
 \langle\chi(0)\chi(m l)\rangle - \langle \chi \rangle^2 &=2\,\sigma^{2}=2l^{4}\beta^{-2}, &\ m =0\\
 &=2\,\bar{\sigma}^{2}=\frac{1}{2}\,l^{4}\beta^{-2}, &\ m = \pm 1\\
 &=0, &\ |m| \geq 2\\
 \\
 \langle\epsilon(0)\epsilon(m\,l)\rangle&=\frac{1}{2}\beta^{-1}, &\ m =0\\
 &=\frac{1}{4}\beta^{-1}, &\ m = \pm 1\\
 &=0, &\ |m| \geq 2
\end{align*}
As expected the correlation functions, being symmetric, depend only on the magnitude and not on the direction of $\bar{x}_{0}=ml$.

Summarizing, the correlation functions for $\chi$ and $\epsilon$ in
$d=1$ are always short ranged, vanishing identically beyond
the nearest neighbor. Correlation functions in the two dimensional
lattice have much more structure as we will see next.

\subsection{The two dimensional harmonic triangular net}

\begin{figure}[h]
\begin{center}
\includegraphics[width=9.0cm]{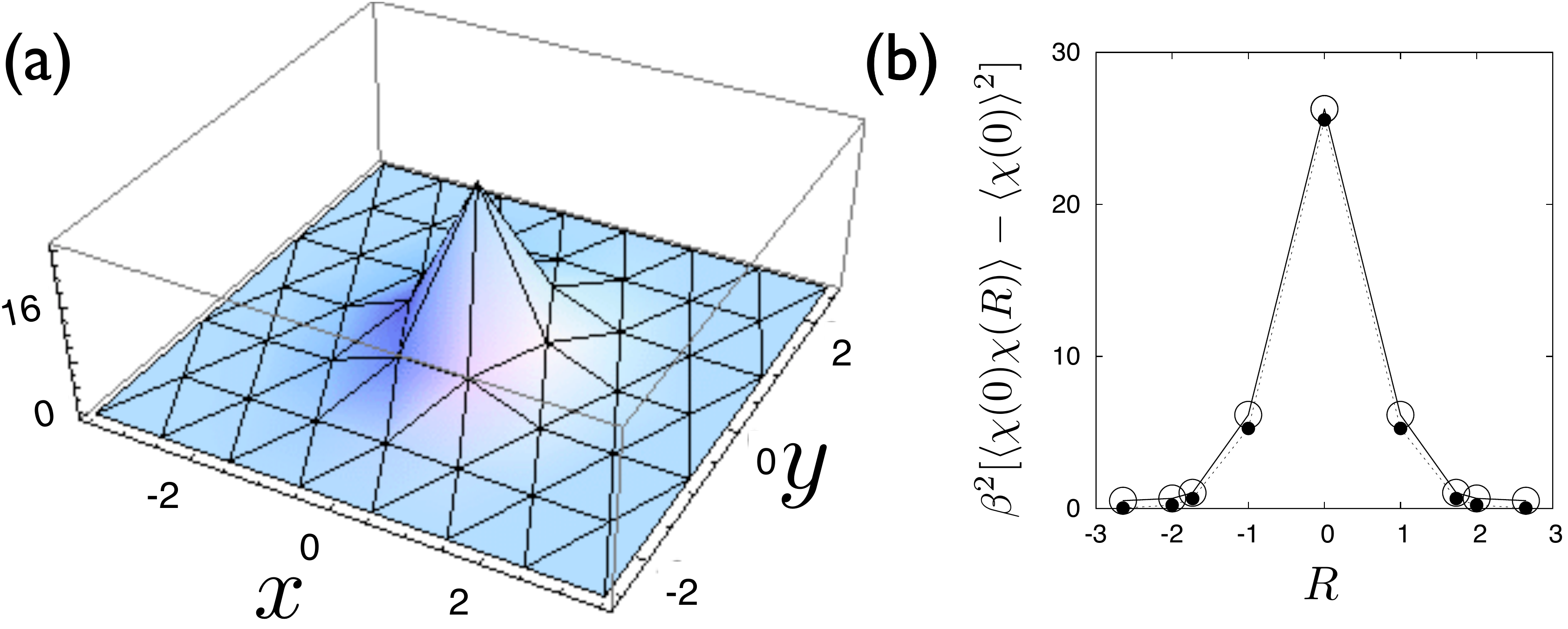}
\end{center}
\caption{(a) Surface plot of $\beta^2(\langle \chi({\bf 0}) \chi({\bf R})
  \rangle - \langle \chi \rangle^2)$ over the two dimensional triangular lattice from our exact
  calculation.  (b) Quantitative comparison for the orientation
  averaged $\beta^2(\langle \chi(0) \chi(R)
  \rangle-\langle\chi\rangle^2)$, where $R = \vert \barR_0 - \bR_0 \vert$. Open circles show the values obtained from our
  exact calculation and filled circles those from simulations of the
  $100\times100$ particle system. The solid lines are guides to the
  eye.}
\label{Fig.4}
\end{figure}

The calculation of the correlation function for the two dimensional
triangular lattice follows along lines similar to that of the single
point distribution functions, except that results now have to be
obtained for each lattice position $\barR_0$. For the $\chi$
correlations, we calculate the trace of $({\mathsf P}\bar{\mathsf C}{\mathsf P})
({\mathsf P}\bar{\mathsf C}{\mathsf P})^t$ as the sum of the $12$ eigenvalues $\bar \sigma_j^2$
and use (\ref{chicor})
to find $\langle \chi({\bf 0}) \chi({\bf R}) \rangle -
\langle\chi\rangle^2$, where $\bR=\barR_0-\bR_0$.
%
%
%
The results obtained using this calculation are compared with
simulations in Fig.~\ref{Fig.4}, showing that this function is
isotropic and decays within about $2-3$ lattice spacings, similar to
the results obtained in Refs.~\cite{kers1,kers2}. 

\begin{figure}[h]
\begin{center}
\includegraphics[width=9.0cm]{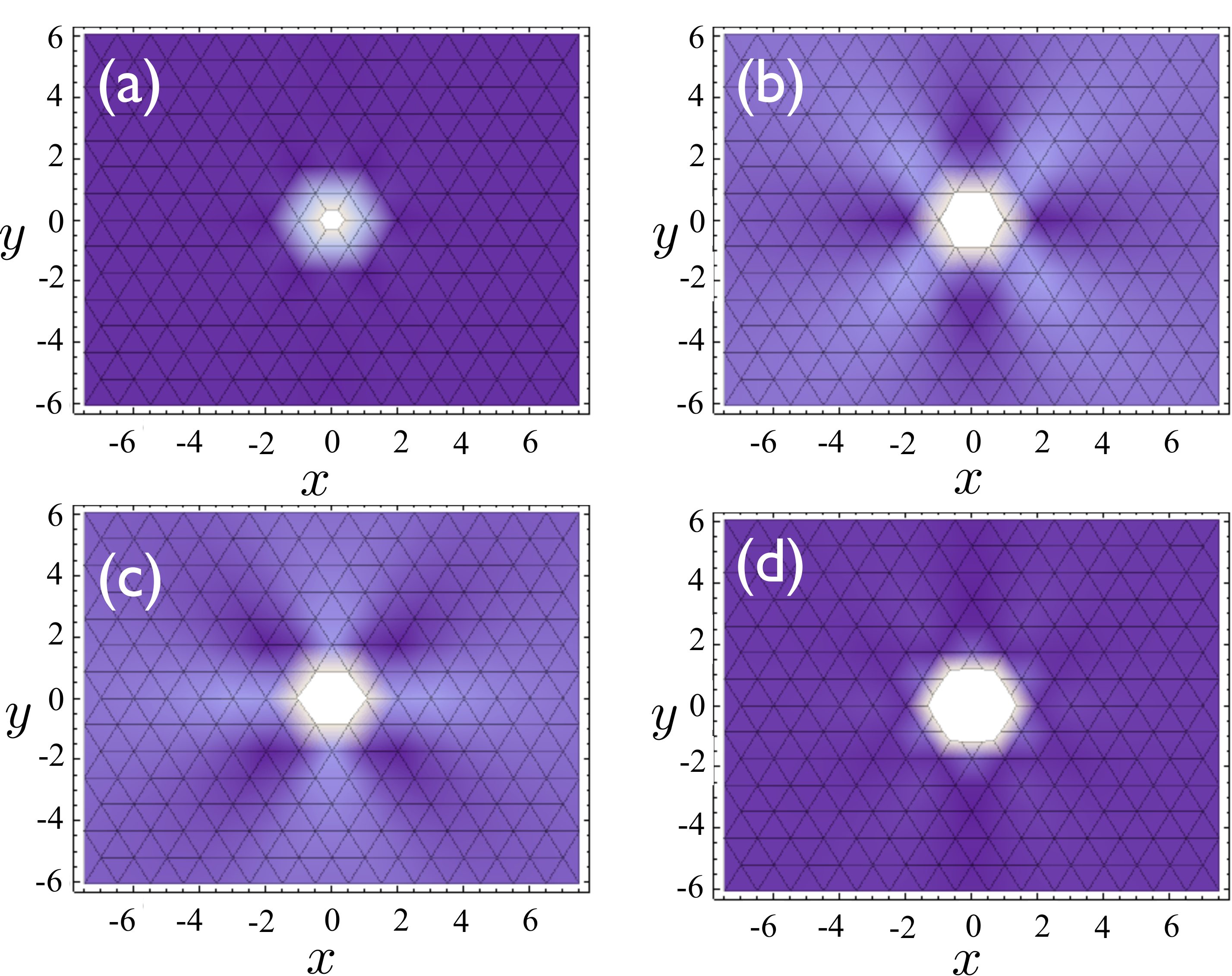}
\end{center}
\caption{(color-online) Density plot of the correlation functions as
  obtained from our calculations showing the shape of the correlation
  functions in the two dimensional plane (a) $\beta\langle e_v({\bf 0})
  e_v({\bf R})\rangle$, (b) $\beta\langle \ed({\bf 0}) \ed({\bf R})\rangle$,
  (c) $\beta\langle e_s({\bf 0}) e_s({\bf R})\rangle$, (d) $\beta\langle
  \omega({\bf 0}) \omega({\bf R}) \rangle$. Note that the volume and rotational correlations are nearly
isotropic while the uniaxial and shear strain correlations show four-fold
anisotropy. The colors vary from dark blue ($-0.1$) to white ($0.2$)
for all the plots. To keep a uniform scale for all graphs we have cut off large values at the origin where necessary. }
\label{Fig.5}
\end{figure}
\begin{figure}[h]
\begin{center}
\includegraphics[width=9cm]{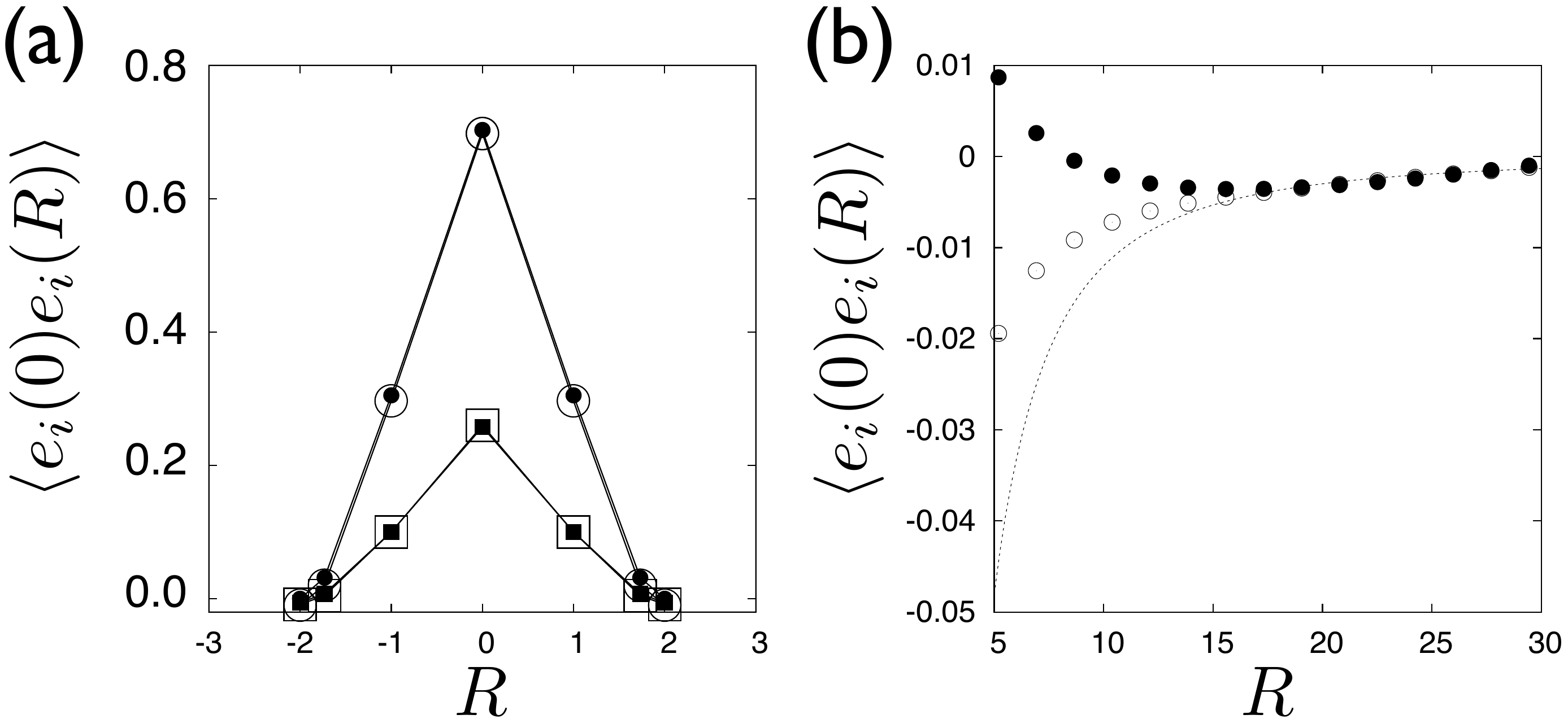}
\end{center}
\caption{(a) Quantitative comparison for the (orientation averaged)
  volume and rotational deformation correlations $ \beta\langle e_v(0)
e_v(R) \rangle $ (squares) and $ \beta\langle \omega (0) \omega (R)
\rangle$ (circles). The open symbols are from our exact calculations
and the filled ones are from our simulations of the $100\times100$
particle system. The solid lines are guides to the eye. The agreement
in the numerical values for the correlations for other components of
the distortion tensor are similar. (b) The large length scale
behaviour of $\beta\langle \ed (0,0) \ed (R/\sqrt{2},R/\sqrt{2})
\rangle$ (open circles) and $\beta\langle e_s (0,0) e_s(0,R) \rangle$
(filled circles) showing the slow decay along these directions. The
dotted  line shows the $\sim R^{-2}$ behaviour for comparison.}
\label{Fig.6}
\end{figure}

Similarly the spatial correlation function for the strain may be
obtained by evaluating $\langle {\bf e} \bar {\bf e}^t\rangle =
\langle {\mathsf Q} {\bf \Delta} \bar {\bf \Delta}^t  {\mathsf Q}^t
\rangle =  {\mathsf Q} \bar {\mathsf C} {\mathsf Q}^t$ for a range of
spatial separations $\bR$.
%
%
The results are shown in Fig.~\ref{Fig.5}. The correlation functions
of $e_v$ and $\omega$ decay rapidly to zero and are nearly isotropic.
We take advantage of this approximate symmetry by averaging  over all pairs $({\bf R}_0,\bar {\bf R}_0)$ related by
symmetry to produce angle-averaged correlation functions that are
functions of $R=|\barR_0 - \bR_0|$ alone.
Results for these functions are compared with those obtained from simulations in Fig.~\ref{Fig.6}(a). 
Correlations involving the uniaxial ($\ed$) and shear ($e_s$) strains
exhibit a pronounced $4$-fold anisotropy at large distances, with
prominent lobes at $0$, $\pi$, $\pm \pi/2$, $\pm \pi/4$ and $\pm 3
\pi/4$.
Furthermore, $\langle \ed ({\bf 0}) \ed({\bf R}) \rangle$ and $\langle
e_s({\bf 0}) e_s({\bf R}) \rangle$ appear to be rotated by $\pi/4$ with
respect to one another (see Fig.~\ref{Fig.5}(b,c)) for large $|{\bf
  R}|$. At small distances, of course, this correspondence is not
exactly satisfied because the underlying triangular lattice does not
have this $\pi/4$ rotational symmetry. 

\begin{figure}[h]
\begin{center}
\includegraphics[width=9.0cm]{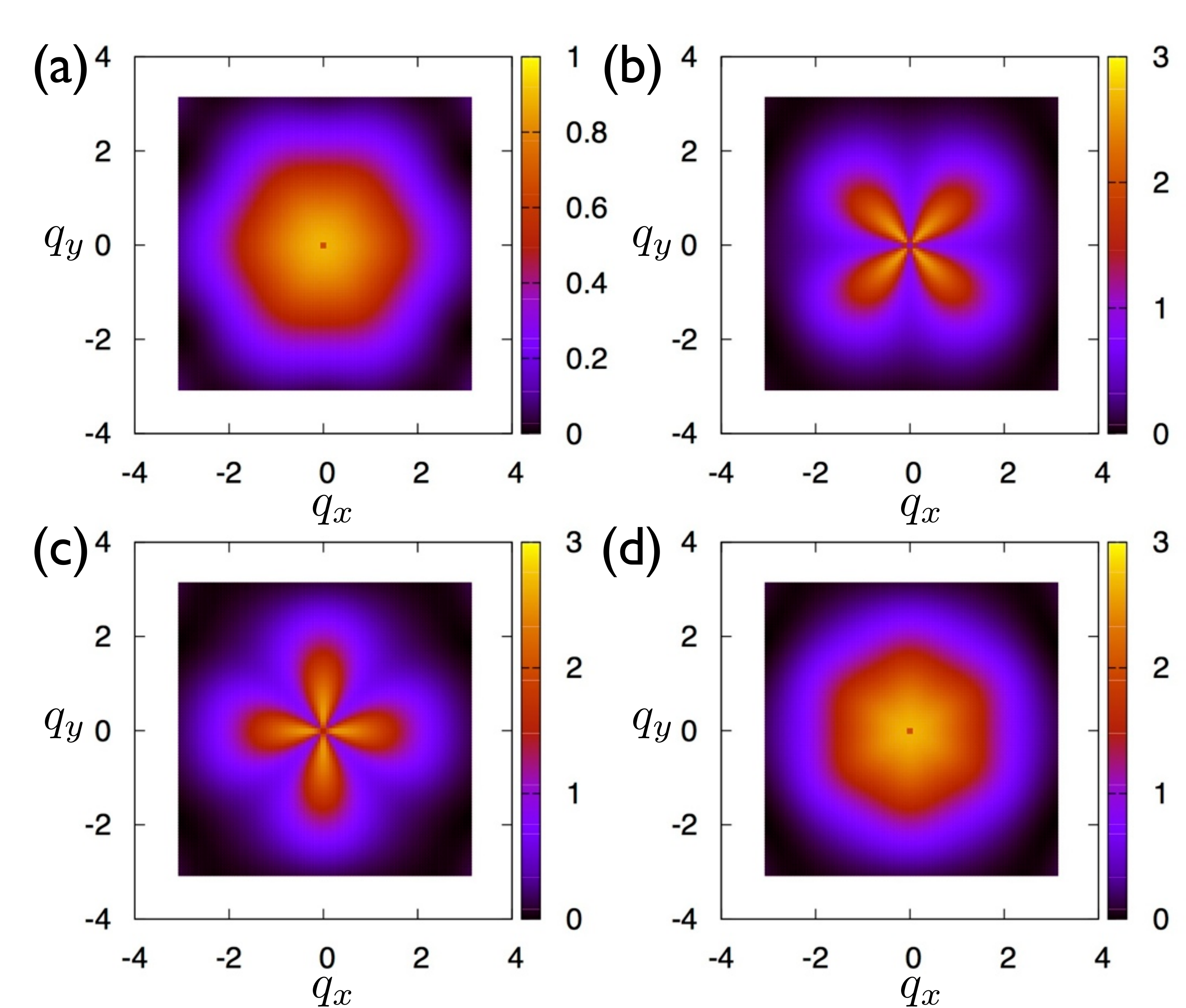}
\end{center}
\caption{(color-online) Plot of the correlation functions in {\bf q}- space obtained from (\ref{cor-q-space}). (a) $\langle e_v^2\rangle({\bf q})$, (b) $\langle \ed^2\rangle({\bf q})$, (c) $\langle e_s^2\rangle({\bf q})$, (d) $\langle \omega^2\rangle({\bf q})$. The key to the colour values are given beside each plot. Note the 4-fold symmetry of the uniaxial and shear correlations similar to the real space plots shown in Fig.~\ref{Fig.5}.}
\label{qcorr}
\end{figure}

An identical $\pi/4$ anisotropy is also observed in momentum space
(Fig.~\ref{qcorr}) where one can obtain closed form expressions for
the correlation functions, in particular in the ${\bf q}\to 0$ limit
as we show below. We begin by writing down explicit expressions for
the strain correlation functions in component form. With the
abbreviation $\Emat = {\mathsf Q}\bar{{\mathsf C}}({\bf q}){\mathsf
  Q}^t$ one has
\begin{eqnarray}
\langle e_v^2\rangle ({\bf q})& = & \Emat_{1111}+\Emat_{2222}+2\Emat_{1122} \nonumber \\
\langle \ed^2\rangle ({\bf q})& = & \Emat_{1111}+\Emat_{2222}-2\Emat_{1122} \nonumber \\
\langle e_s^2\rangle ({\bf q})& = & \Emat_{1212}+\Emat_{2121}+2\Emat_{1221} \nonumber \\
\langle \omega^2\rangle ({\bf q})& = & \Emat_{1212}+\Emat_{2121}-2\Emat_{1221}
\label{q-corr-array}
\end{eqnarray}
The notation used here is the same as in (\ref{strain_FT_general}),
so that e.g.\ $\langle e_v^2\rangle ({\bf q})$ is the Fourier transform of the
strain correlator $\langle e_v\bar{e}_v\rangle \equiv \langle
e_v(\bR_0)e_v(\barR_0)\rangle = \vBZ^{-1} \int d{\bf q}\,
\langle e_v^2({\bf q})\rangle e^{i{\bf q}\cdot(\bR_0-\barR_0)}$. 
After substituting in for $\bar{\mathsf C}({\bf q})$ from
(\ref{cor-q-space}) one gets, using also that for the triangular
lattice ${\mathsf R}^t{\mathsf R}$ is three times the identity matrix,
\begin{eqnarray}
\Emat_{\alpha\alpha'\gamma\gamma'} & = & \frac{1}{9}\sum_{i,j = 1}^6
f^{ij} R_{i\alpha'} R_{j\gamma'}\Dmat^{-1}_{\alpha\gamma}({\bf q})
\label{qcq}
\end{eqnarray}
with
\begin{eqnarray}
f^{ij} & = & f^{ij}_R + i f^{ij}_I \\
f^{ij}_R & = & \cos[{\bf q}\cdot({\bf R}_i - {\bf R}_j)] - \cos({\bf q}\cdot{\bf R}_i) - \cos({\bf q}\cdot{\bf R}_j)+1 \nonumber \\
f^{ij}_I & = & \sin[{\bf q}\cdot({\bf R}_m - {\bf R}_n)] - \sin({\bf q}\cdot{\bf R}_i) + \sin({\bf q}\cdot{\bf R}_j)\nonumber
\end{eqnarray}
Noting that the imaginary contributions to $\Emat$ sum to zero and
expanding $f^{ij}_R$ in Eqn.~(\ref{qcq}) yields for small $q$
\begin{eqnarray}
\Emat_{\alpha\alpha'\gamma\gamma'} & = & 
 \frac{1}{9}\sum_{i,j = 1}^6 \sum_{\mu\nu}
q_\mu q_\nu R_{i\mu} R_{j\nu} R_{i\alpha'} R_{j\gamma'}\Dmat^{-1}_{\alpha\gamma}({\bf
  q})
\nonumber \\
& = & q_{\alpha'} q_{\gamma'}\Dmat^{-1}_{\alpha\gamma}({\bf q})
\label{qcq2}
\end{eqnarray}
where we have used again that $\sum_i R_{i\alpha}
R_{i\gamma}=3\delta_{\alpha\gamma}$. 
To examine the leading order behaviour at small ${\bf q}$ for the
strain correlators, we expand ${\mathcal
  D}^{-1}({\bf q})$ about ${\bf q} = 0$ in (\ref{qcq2}) and substitute
into (\ref{q-corr-array}) to obtain
\begin{eqnarray}
\langle e_v^2\rangle ({\bf q})& \approx & \frac{8}{9}
\\
\langle \ed^2\rangle ({\bf q})& \approx & \frac{8}{9} + \frac{64}{9}\frac{q_x^2\,q_y^2}{(q_x^2 + q_y^2)^2} \label{eu_q}\\
\langle e_s^2\rangle ({\bf q})& \approx & \frac{8}{3} - \frac{64}{9}\frac{q_x^2\,q_y^2}{(q_x^2 + q_y^2)^2} \label{es_q} \\
\langle \omega^2\rangle ({\bf q})& \approx & \frac{8}{3}
\label{q-corr-array2}
\end{eqnarray}
consistent with the results plotted in Fig.~\ref{qcorr}. The fact that
the correlators of uniaxial strain $\ed$ and shear strain $e_s$ are
related by a rotation becomes clearer if one rewrites
(\ref{es_q}) as
\begin{eqnarray}
\langle e_s^2({\bf q})\rangle & \approx & \frac{8}{9} +
\frac{64}{9(q_x^2 + q_y^2)^2}\left[
\left(\frac{q_x+q_y}{\sqrt{2}}\right)^2
\left(\frac{q_x-q_y}{\sqrt{2}}\right)^2
\right] \nonumber \\
\label{es_q_again}
\end{eqnarray}
This evidently maps to (\ref{eu_q}) under a rotation by $\pi/4$, as claimed.
We note that the large distance (small {\bf q}) anisotropies of
$\langle \ed^2({\bf q})\rangle$ and $\langle e_s^2({\bf q})\rangle$
are also consistent with a mean field, continuum theory, calculation
shown in detail in Ref.~\onlinecite{kers2}. Interestingly, that
calculation restricted attention to fluctuating strain fields that
satisfy force balance, so one concludes that it is indeed these
configurations that dominate the scaling for large distances.

The ${\bf q}$ space structure of the uniaxial and the shear
correlation functions as implied by (\ref{q-corr-array2}),
Figs.~\ref{qcorr}(b) and (c), and further supported by continuum
theory~\cite{kers2}, shows that these correlation functions have
singularities at $q = 0$, with the second terms in (\ref{eu_q}) and
(\ref{es_q_again}) vanishing along specific directions ($q_x = 0, q_y = 0$
in $\ed$ and $q_x = \pm q_y$ in $e_s$). These singularities lead to slow ($\sim
1/R^2$) decay of the correlation functions in real space (see
Fig.~\ref{Fig.6}(b)), with the prefactor alternating in sign
according to $\cos(4\theta)$ or $\sin(4\theta)$ with the polar angle of $\bR$.

Before we end this section, we remark that the strain correlation functions, by linear response, are proportional to the strain field produced by a point, delta function stress at the origin. Since the large $R$ (or small $q$) structure of the correlation functions is insensitive to crystal symmetry, it is no surprise that similar quadrupolar displacement patterns~\cite{quad} have been observed in association with local rearrangements in amorphous materials known as shear transformation zones (STZ)~\cite{argon,falk,falk-review}, in both experiments\cite{schall} and computer simulations~\cite{lemat,barat} of granular or glassy materials under shear. 
  

\section{Linear response and the non-affine field}
\label{sec:4}

The form of the characteristic function (\ref{joint1}) offers a simple
way to calculate the response of the system to uniform fields
conjugate to $\chi$  and ${\bf e}$. We analytically continue $k$ and
$\bkappa$ to the complex plane by replacing $\bkappa \to \bkappa -
i\,{\bf \Sigma}$ and $k \to k - i\,h_{\chi}$. Here the vector ${\bf
  \Sigma}$, once rearranged into a symmetric tensor, is the
stress~\cite{LL}, made dimensionless by multiplying by the inverse
temperature $\beta$ and the size of a suitable local volume of the order of $R_\Omega^d$. On the other hand $h_{\chi}$ is a new field, conjugate to
$\chi$. The introduction of a (small) stress merely shifts $\langle
{\bf e} \rangle$ away from zero to a value proportional to $\Sigma$ (Hooke's law), i.e.\ $\langle
e_i\rangle_{\bf \Sigma} = \langle e_i e_j \rangle_{{\bf
    \Sigma}=0} \Sigma_j$. The proportionality constant here is the zero
field compliance calculated earlier.
Furthermore, because of the coupling between ${\bf e}$ and $\chi$, external stress (mainly shear and uniaxial, see Section\,~\ref{sec:2} and 
Eq.\,(\ref{cross})) will change $P(\chi)$ for lattices where $[{\mathsf P},{\mathsf C}]$ and higher commutators are non-zero.
A straightforward calculation, introducing ${\bf \Sigma}$ in (\ref{full-phi}) and Taylor expanding, shows that for small values of 
${\bf \Sigma}$, 
 $$\langle\chi\rangle_{{\bf \Sigma}}=\langle\chi\rangle_{{\bf \Sigma}=0} + {\bf \Sigma}^{t}{\mathsf Q}{\mathsf C}[{\mathsf P},{\mathsf C}]{\mathsf Q}^{t}{\bf \Sigma},$$
 which always increases $\langle \chi \rangle$ for the $d=2$
 triangular lattice; higher moments of $\chi$ are similarly
 affected. At large ${\bf \Sigma}$, of course, perturbations would
 become so large that the effects of anharmonicities that we have not
 modelled would become apparent.

\begin{figure}[h]
\begin{center}
\includegraphics[width=9.5cm]{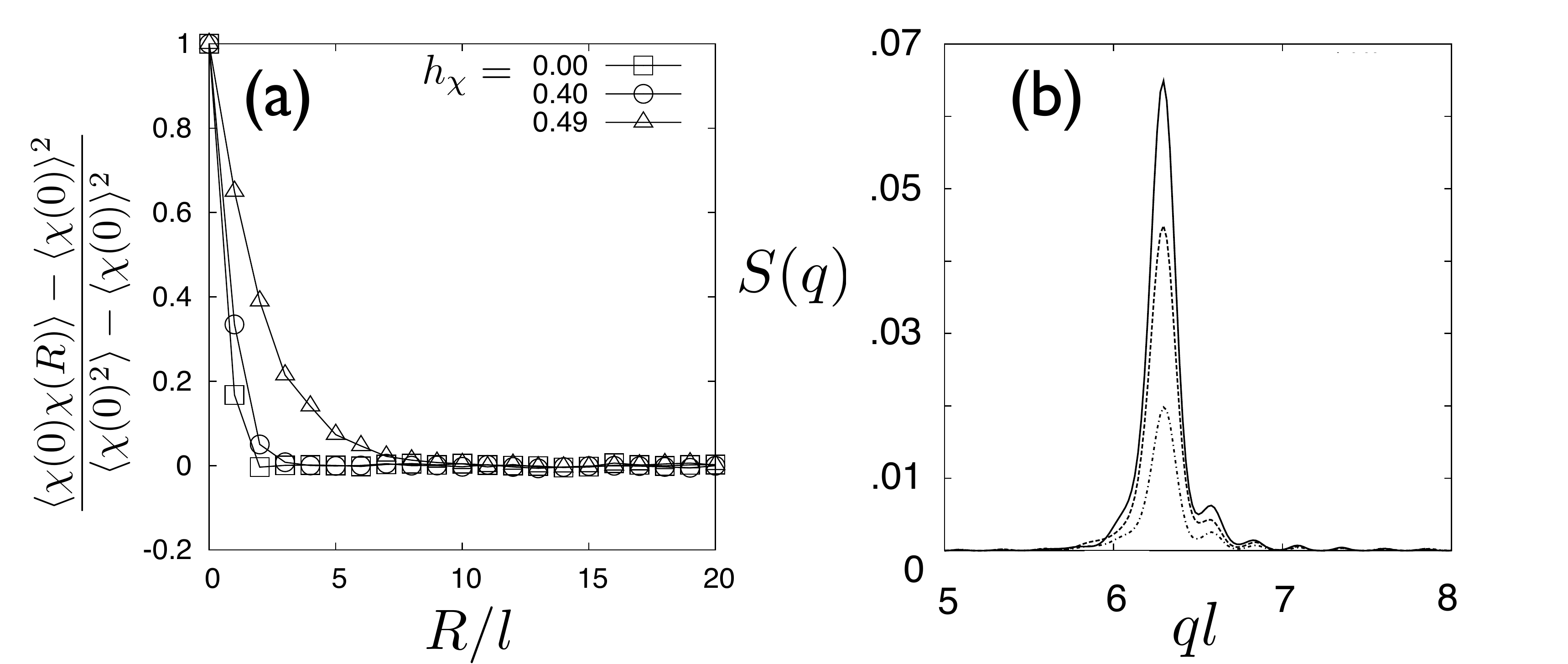}
\end{center}
\caption{(a) The scaled, two point $\chi$-$\chi$ correlation function
  for $h_{\chi} = 0$, $0.4$ and $0.49$ from Monte Carlo simulations
  of a $50$ particle harmonic chain at $\beta = 100$, showing a large
  increase in the correlation length as the critical field is
  approached. 
  (b) The first peak in the structure factor $S(q) = \langle \rho_q
  \rho_{-q} \rangle$ for $h_{\chi} = 0$ (solid line), $0.4$ (dashed
  line) and $0.49$ (dash-dotted line) for the same system as in
  (a). Note that the amount of structure is reduced at constant temperature.}
\label{sofq}
\end{figure}

The effect of $h_{\chi}$ is also intriguing. Below we illustrate this
explicitly for the one-dimensional case, though similar results should hold
in any dimension and for particles with arbitrary interactions. The
joint characteristic function for $\chi$ and $e$ for the $d=1$ chain
with $h_\chi$ included is 
\begin{align}
\varPhi(k,\kappa)&=\sqrt{\frac{1-2\,\sigma\,h_{\chi}}{1-2\,\sigma\,ik-2\,\sigma\,h_{\chi}}}
\exp\left(-\frac{1}{2}\langle\epsilon^{2}\rangle\kappa^{2}\right)
\label{Phi_reweighted}
\end{align}
Note that we have multiplied $\varPhi(k,\kappa)$ by the factor $\sqrt{1-2\,\sigma\,h_{\chi}}$ to ensure normalization, i.e.\
$\text{lim}_{k\rightarrow0,\kappa\rightarrow0} \varPhi(k,\kappa)= 1$. 
Now we can obtain $P(\epsilon)$, $P(\chi)$ and $P(\chi,\epsilon)$ by
inspection as
\begin{eqnarray}
P(\chi,\epsilon) & = & P(\chi)  P(\epsilon)\nonumber \\
& = & \sqrt{\frac{1-2\,\sigma\,h_{\chi}}{2\pi\,\sigma}}\chi^{-1/2}\exp\left[-\frac{(1-2\,\sigma\,h_{\chi})\chi}{2\,\sigma}\right] \times \nonumber \\
& & \frac{1}{\sqrt{2\pi\langle\epsilon^{2}\rangle}}\exp\left[-\frac{\epsilon^{2}}{2\langle\epsilon^{2}\rangle}\right]
%
\end{eqnarray}
As $h_{\chi}\rightarrow 1/(2\sigma)$, $P(\chi)$ becomes proportional
to $\chi^{-1/2}$ and all the moments 
$$\langle\chi^{n}\rangle  = \left(\frac{2\sigma}{1-2\,\sigma\,h_{\chi}}\right)^{n}\frac{\Gamma(n+\frac{1}{2})}{\Gamma(\frac{1}{2})} $$
diverge as $[h_{\chi}-1/(2\sigma)]^{-n}$. Spatial correlations of
non-affinity $\chi$ also become long ranged in this limit and the
system becomes disordered. Displacements acquire a non-affine
character over all length scales, as evidenced from the decrease of the amplitude of the structure factor for a finite one-dimensional chain of $100$ particles, see Fig.~\ref{sofq}(a) and (b).

All of this suggests the presence of a critical line in the $(h_{\chi},\beta)$ plane beyond which the system becomes globally non-affine so that $\chi$ defined by coarse graining over local neighborhoods
$\Omega$ is always infinite -- a ``maximally non-affine" solid (the presence of anharmonicities would, in practice, be expected to limit $\chi$ to a finite value).  For
the $d=1$ lattice, this transition is identical to the celebrated
Peierls transition~\cite{ashcroft} as can be deduced from the nature
of the non-affine mode shown in Fig.~\ref{1dlat}(c). In higher
dimensions, the transition appears for values of $h_{\chi}$ equal to
a critical $h_{\chi}^*$ close to half the reciprocal of the largest eigenvalue of ${\mathsf P}{\mathsf
  C}{\mathsf P}^t$. Indeed, one finds easily by writing down the general analog of
(\ref{Phi_reweighted}), taking a log and differentiating w.r.t.\ $ik$
that
\begin{eqnarray}
\langle\chi\rangle & = & {\rm Tr}(\mathsf{I}-2h_\chi {\mathsf P}{\mathsf
  C}{\mathsf P}^t)^{-1}
\end{eqnarray}
which diverges at $h_\chi^* = 1/(2\max_j \sigma_j)$ if the $\sigma_j$
are the eigenvalues of ${\mathsf P}{\mathsf C}{\mathsf P}^t$ as before.
In the triangular net, the largest eigenvalues are $\sigma_1 =
\sigma_2$ and the displacement patterns whose amplitudes would grow at
the transition are shown in Fig.~\ref{Fig.7}(a) and (b).  Locally, they
correspond to an almost uniform translation of the neighbouring particles relative to the central particle, though it is difficult to visualise what global configuration is finally produced. We imagine that this leads to destruction of the lattice structure and eventual amorphisation. Of course, as the transition is approached, we expect the identity of the neighbourhoods, $\Omega$, itself to become ill-defined due to this loss of crystalline order, making many of our results invalid in that limit.
 
In the harmonic lattice, the transition discussed above is {\em
  hidden} because the only physically realisable value of the
non-affine field $h_{\chi} = 0$ lies on the critical line at infinite
temperature $\beta = 0$. Nevertheless, there may be systems where the
critical line cuts the $h_{\chi} = 0$ axis at a non-zero value of
$\beta$ (i.e.\ at finite temperature). In such a system one may obtain a
physical transition from an affine to a maximally non-affine state as
the temperature is increased or a stiffness parameter is reduced producing a going
over to  a ``glass spinodal''~\cite{pia}. We speculate that this may
also happen at or near the yield point of a solid under external load
where the bonds between atoms become weak due to strong
anharmonicities. However, for such cases, the simple linear response
calculations presented above becomes invalid and the distributions of
$\chi$ and ${\bf  e}$ become strongly coupled through the strain and
non-affinity dependence of the dynamical matrix ${\Dmat}({\bf
q})$. Implications of this transition for the mechanical and phase
behaviour of solids in $1$, $2$ and $3$ dimensions are being worked out
and will be published elsewhere.

\section{Summary and Conclusions}
\label{sec:5}

In this paper we have shown that coarse graining of the microscopic
displacements of crystalline solid at non-zero temperatures generates
non-affine as well as locally affine distortions. The procedure effectively
amounts to integrating out phonon modes with wavelengths comparable to
or smaller than the coarse-graining length. We have obtained the
probability distributions for the non-affine parameter and the affine
distortions of a harmonic lattice at non-zero temperatures. We have
also obtained the spatial correlations of the local distortions and
non-affinity. While $\langle \chi({\bf 0}) \chi(\bR) \rangle$ decays
exponentially, the correlation functions corresponding to the
different elements of the distortion tensor decay differently. Volume
and rotation correlations, on the one hand, are short ranged; uniaxial
and shear strain components, on the other hand, decay as a $1/R^2$ power law
with prefactor depending on the polar angle as $\sin(4\theta)$ or
$\cos(4\theta)$, respectively. The angular dependences
of the slow decay for uniaxial strain and shear are rotated by $\pi/4$
with respect to each other. We noted that this is consistent with an earlier
continuum calculation imposing the mechanical stability condition that
the stress is divergence free in the fluctuating strain field.
Finally, we have shown
that it is possible to induce a transition from a solid with a finite
average non-affinity $\langle \chi \rangle$ to one where all moments $\langle \chi^n \rangle$ diverge, by tuning a non-affine field $h_{\chi}$. 

\begin{figure}[h]
\begin{center}
\includegraphics[width=6cm]{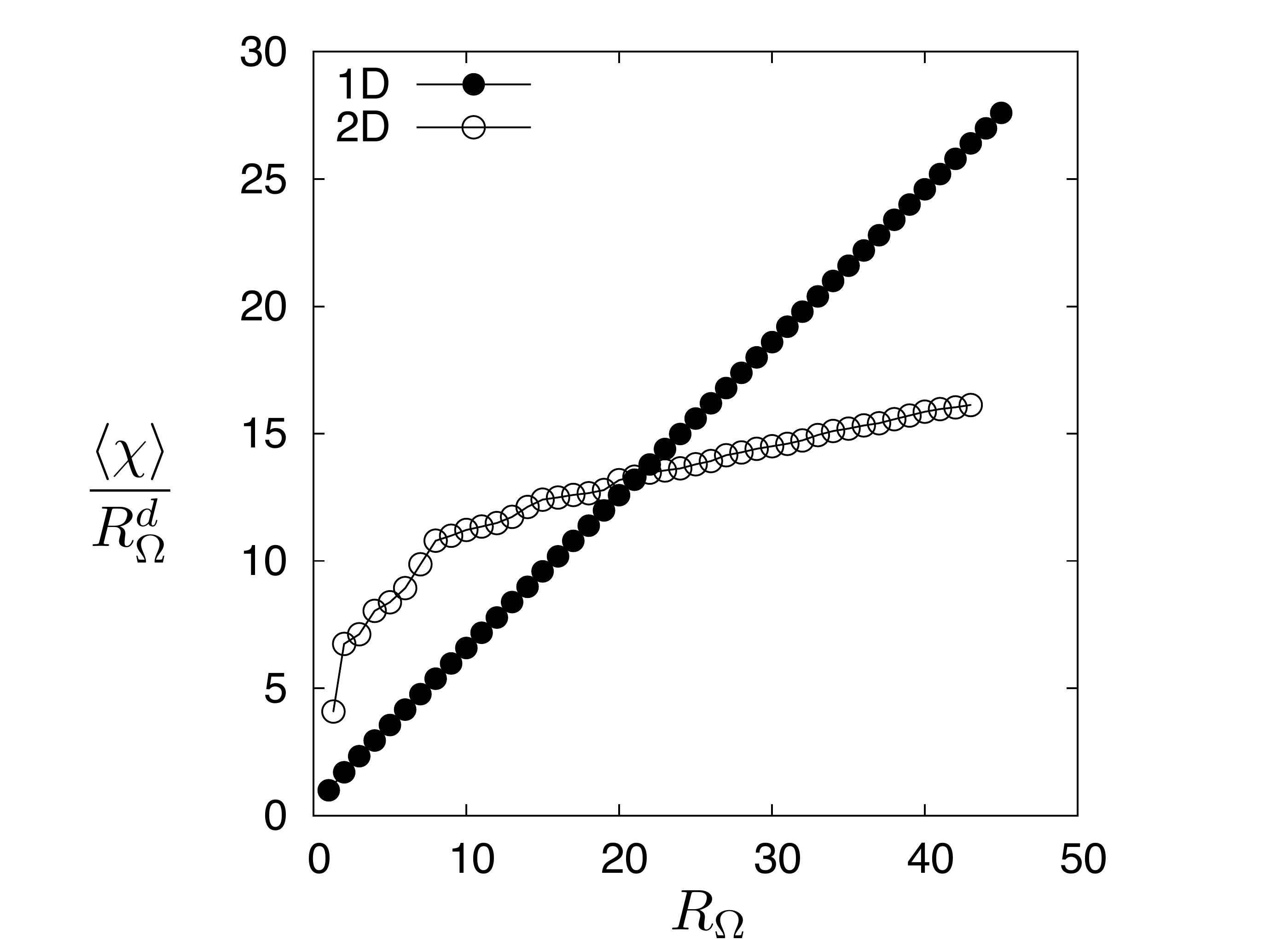}
\end{center}
\caption{Plot of $\langle \chi \rangle /R^d_{\Omega}$ vs $R_{\Omega}$
  for $d=1$ (filled circles) and $d=2$ (open circles). Note that while
  the intensive (per unit coarse-graining volume) $\chi$ increases
  linearly with $R_\Omega$, the size of the reference volume $\Omega$,
  in $d=1$, it has a much slower logarithmic increase in $d=2$.
}
\label{FSS}
\end{figure}

How do our results vary with the size of the reference volume $\Omega$
used to define the coarse-grained quantities? Firstly, $\chi$ depends
on the number of sites within $\Omega$, so we need to consider the
intensive quantity $\langle \chi \rangle /R^d_{\Omega}$ where
$R_{\Omega}$ is the radius of the reference volume
$\Omega$. Furthermore, since $\chi$ depends on the fluctuations of the
displacements ${\bf u}$, then in $d=1$ this quantity itself should
diverge as $\sim R_{\Omega}$ and in $d=2$ as $\sim
\log(R_{\Omega})$~\cite{CL}. In Fig.~\ref{FSS} we show plots for this
normalized non-affinity in $d=1$ and $d=2$, which confirm these expectations.
In higher dimensions the intensive variable $\langle \chi \rangle
/R^d_{\Omega}$ should for large $R_\Omega$ approach a constant (proportional to temperature, as for $d=1$ and 2). Note that the probability distribution of strains,
$P(\chi,{\bf e}_{j})$, is similarly $\Omega$ dependent and so our
calculation automatically incorporates exact finite size scaling of
the elastic compliances. Approximate finite size scaling results based
on continuum elasticity theory have been used to obtain elastic
constants of colloidal solids from video microscopy
data~\cite{zahn,kers1}. Our results may offer a better way to analyze
such data.

Secondly, as $R_{\Omega}$ increases, $\chi$ and the distortion
${\mathsf D}$ may get more and more coupled. This is best illustrated,
for the $d=1$ harmonic chain, by computing the norms of the
commutators $[{\mathsf P},{\mathsf C}]$ and $[{\mathsf C}{\mathsf
  P},[{\mathsf P},{\mathsf C}]]$ which we obtain as shown in
Table~\ref{R_omega_table}.
\begin{table}[h!]
\centering
\begin{tabular}{l c c r}
$R_{\Omega}/l$ & $\lVert{\mathsf C}\rVert$ & $\lVert[{\mathsf P},{\mathsf C}]\rVert$ & $\lVert[{\mathsf C}{\mathsf P},[{\mathsf P},{\mathsf C}]]\rVert$ \\
$1$ & $1.414$ & $0.000$ &  $0.000$\\    
$2$ & $3.741$ & $0.282$ & $0.126$\\   
$3$ & $7.211$ & $0.654$ & $0.534$\\
$4$ & $11.832$ & $1.148$ & $1.516$\\ 
$5$ & $17.606$ & $1.766$ & $3.448$\\
$6$ & $24.535$ & $2.507$ & $6.803$\\
$7$ & $32.619$ & $3.371$ & $12.145$\\
$8$ & $41.856$ & $4.358$ & $20.132$\\
$9$ & $52.249$ & $5.469$ & $31.520$\\
$10$ & $63.796$ & $6.704$ & $47.154$\\
\end{tabular}
\caption{Numerical values of the norm of the successive commutators of
  ${\mathsf P}$ and ${\mathsf C}$ which contribute to the cross
  correlation of non-affinity and strain (see
  section\,~\ref{sec:1}). Note that the commutators increase with the
  size $R_{\Omega}$ of the reference volume. }
\label{R_omega_table}
\end{table}

For very large $R_{\Omega}$ more and more terms are needed to get good convergence of the Taylor expansion for $P(\chi,{\bf e}_{j})$ in powers of $k$,
viz.\ Eq.~(\ref{taylor}),  for given $k$. Accordingly $\chi$ becomes inextricably linked with the strain, a phenomenon related to the fluctuation-driven instability of ordered solids in one dimension~\cite{CL}. This effect should be weaker in higher dimensions, though a full study of the influence of dimensionality on non-affinity for a variety of lattices and interactions needs further work.  


Our calculations may be easily extended to other lattices and to
higher dimensions without much difficulty, requiring at most a
calculation of the relevant dynamical matrix. Similarly, local $\chi$
and displacement distributions can be obtained for crystals with
isolated, point or line defects once the appropriate Hessians of the
local potential energy at defects are evaluated. The effect of
external stress on $\chi$ is another interesting problem which may be
addressed immediately in the limit of negligible anharmonicity; the
effect of anharmonic terms could be included perturbatively. Finally the effect of disorder can be incorporated~\cite{zero-T} at non-zero temperatures. 

Our results may also be used to construct new simulation strategies
for investigating the mechanical behaviour of solids under external
stress. For instance, particle moves may be designed which change
$\chi$ without influencing the local distortion within $\Omega$ by
projecting particle displacements along eigendirections of ${\mathsf
  P}$.

Such calculations will be particularly useful for glasses~\cite{falk-review}, where local potential energy Hessians may be used to define an equivalent harmonic lattice at every instance of time with $\chi$ being calculated dynamically
from the reference configurations at the previous timestep.
Calculations similar to ours will be useful to understand the
properties of shear transformation zones (STZ)\,\cite{argon}, defined~\cite{falk, falk-review} as
regions with a large value of $\chi$; these are the dominant entities
responsible for mechanical deformation of glasses. STZ are thermally
generated and respond to external stress by rearrangements of local
particle positions. Similar localised non-affine excitations have also been
observed in anharmonic, crystalline solids~\cite{tommy} where they
have been identified as droplet fluctuations from nearby glassy and
liquid-like minima of the free energy. Constrained simulations like
those outlined above may help in identifying the role of $\chi$ in the
processes involved in complex phenomena like anelasticity, yielding
and melting. The role of the non-affine field $h_{\chi}$ in
influencing glass transition and the mechanical behaviour of solids
both crystalline and amorphous is another direction that we intend to investigate in the future.  
  

\acknowledgments
SG and SS thank the DST India (Grant No.INT/EC/MONAMI/(28)233513/2008) for support. The present work was formulated during a visit by some of the authors (SS, PS and MR) to the KITP Santa Barbara, where this research was supported in part by the National Science Foundation under Grant No. NSF PHY05-51164. Discussions with Srikanth Sastry are gratefully acknowledged.



\begin{thebibliography}{}
\bibitem{metric} Our definition of $\Omega$, implicitly assumes that fluctuations do not change the number and identity of the particles within the neighbourhood since we study a network with a fixed topology of connections.
\end{thebibliography}
\end{document}